\documentclass[aps,prb,reprint,preprintnumbers,amsmath,amssymb,showpacs,superscriptaddress,citeautoscript,longbibliography]{revtex4-2}
\usepackage{amssymb}

\usepackage{graphicx}% Needed to include png/jpg/pdf figure files

\usepackage{amsmath}
\usepackage{bm}
\usepackage{color, soul}
\usepackage[normalem]{ulem}
\usepackage{siunitx}
\usepackage{upgreek}
\usepackage[dvipsnames]{xcolor}
\usepackage{comment}
\usepackage{natbib}
\usepackage{braket}
\usepackage{ulem}

\usepackage[breaklinks]{hyperref}%hyperef package does internal and external linking (labels, urls)
\usepackage[all]{hypcap}%addition to hyperref, it ensures figure is correctly displayed (and not only the caption) when link is clicked
\hypersetup{
    plainpages=false,
    unicode=false,          % non-Latin characters in AcrobatĂŻÂżÂ˝s bookmarks
    pdfmenubar=true,        % show AcrobatĂŻÂżÂ˝s menu?https://www.overleaf.com/13403665mnpkhvmzmbnjhttps://www.overleaf.com/13403665mnpkhvmzmbnj
    pdffitwindow=false,     % window fit to page when opened
    pdfstartview={FitH},    % fits the width of the page to the window
    pdftitle={CrSBr_Chlorine},    % title
    pdfauthor={Maciej Śmiertka},     % author
    pdfproducer={PWr}, % producer of the document
    pdfkeywords={High} {Magnetic} {Fields}, % list of keywords
    pdfnewwindow=true,      % links in new window
    linktoc=section,
    colorlinks=true,       % false: boxed links; true: colored links
    linkcolor=blue,          % color of internal links
    citecolor=red,        % color of links to bibliography
    filecolor=magenta,      % color of file links
    urlcolor=blue           % color of external links
}

\DeclareUnicodeCharacter{2212}{-}

\usepackage{xr}
\begin{document}

\title{Tunable Magneto-Excitonic Coupling in Alloyed van der Waals Antiferromagnet}

%Tuning the Wannier-Mott to Frenkel character of excitons through halogen substitution in CrSBr
%Chemical tuning of the Wannier-Mott to Frenkel character of excitons in CrSBr

\author{Maciej Śmiertka}
\affiliation{Department of Experimental Physics, Faculty of Fundamental Problems of Technology, Wroclaw University of Science and Technology, 50-370 Wroclaw, Poland}

\author{Oliwia Janikowska}
\affiliation{Department of Experimental Physics, Faculty of Fundamental Problems of Technology, Wroclaw University of Science and Technology, 50-370 Wroclaw, Poland}

\author{Katarzyna Olkowska-Pucko}
\affiliation{Faculty of Physics, University of Warsaw, 02-093 Warsaw, Poland}

\author{Grzegorz Krasucki}
\affiliation{Faculty of Physics, University of Warsaw, 02-093 Warsaw, Poland}

\author{Katarzyna Posmyk}
\affiliation{Department of Experimental Physics, Faculty of Fundamental Problems of Technology, Wroclaw University of Science and Technology, 50-370 Wroclaw, Poland}
\affiliation{Laboratoire National des Champs Magn\'etiques Intenses, EMFL, CNRS UPR 3228, Universit{\'e} Grenoble Alpes, Universit{\'e} Toulouse, Universit{\'e} Toulouse 3, INSA-T, Grenoble and Toulouse, France}

\author{Paulina Peksa}
\affiliation{Department of Experimental Physics, Faculty of Fundamental Problems of Technology, Wroclaw University of Science and Technology, 50-370 Wroclaw, Poland}
\affiliation{Laboratoire National des Champs Magn\'etiques Intenses, EMFL, CNRS UPR 3228, Universit{\'e} Grenoble Alpes, Universit{\'e} Toulouse, Universit{\'e} Toulouse 3, INSA-T, Grenoble and Toulouse, France}
\author{Alessandro Surrente}
\affiliation{Department of Experimental Physics, Faculty of Fundamental Problems of Technology, Wroclaw University of Science and Technology, 50-370 Wroclaw, Poland}

\author{Dimitar Pashov}
\affiliation{King’s College London, Theory and Simulation of Condensed Matter, The Strand, WC2R 2LS London, UK}
\author{Kseniia Mosina}
\affiliation{Department of Inorganic Chemistry, University of Chemistry and Technology Prague, Technicka 5, Prague 6, 16628 Czech Republic}

\author{Zdenek Sofer}
\affiliation{Department of Inorganic Chemistry, University of Chemistry and Technology Prague, Technicka 5, Prague 6, 16628 Czech Republic}

\author{Mark~van Schilfgaarde}
\affiliation{National Laboratory of the Rockies, Golden, CO, USA}

\author{Adam Babiński}
\affiliation{Institute of Experimental Physics, Faculty of Physics, University of Warsaw, Pasteura 5, 02-093, Warsaw, Poland}

\author{Maciej R. Molas}
\affiliation{Institute of Experimental Physics, Faculty of Physics, University of Warsaw, Pasteura 5, 02-093, Warsaw, Poland}

\author{Gabriela Komorowska}
\affiliation{Warsaw University of Technology, Faculty of Materials Science and Engineering, Woloska 141, 02-507 Warsaw, Poland}

\author{Esteban Zamora-Amo}
\affiliation{2D Foundry Research Group, Instituto de Ciencia de Materiales de Madrid (ICMM-CSIC), Madrid E-28049, Spain}

\author{Andres Castellanos-Gomez}
\affiliation{2D Foundry Research Group, Instituto de Ciencia de Materiales de Madrid (ICMM-CSIC), Madrid E-28049, Spain}
\author{Federico Mompeán}
\affiliation{2D Foundry Research Group, Instituto de Ciencia de Materiales de Madrid (ICMM-CSIC), Madrid E-28049, Spain}

\author{Mar Garcia-Hernandez}
\affiliation{2D Foundry Research Group, Instituto de Ciencia de Materiales de Madrid (ICMM-CSIC), Madrid E-28049, Spain}
\author{Micha{\l} Baranowski}\email{michal.baranowski@pwr.edu.pl}
\affiliation{Department of Experimental Physics, Faculty of Fundamental Problems of Technology, Wroclaw University of Science and Technology, 50-370 Wroclaw, Poland}

\author{Swagata~Acharya}\email{swagata.acharya@nlr.gov}
\affiliation{National Laboratory of the Rockies, Golden, CO, USA}

\author{Paulina Plochocka}\email{paulina.plochocka@lncmi.cnrs.fr}
\affiliation{Department of Experimental Physics, Faculty of Fundamental Problems of Technology, Wroclaw University of Science and Technology, 50-370 Wroclaw, Poland}
\affiliation{Laboratoire National des Champs Magn\'etiques Intenses, EMFL, CNRS UPR 3228, Universit{\'e} Grenoble Alpes, Universit{\'e} Toulouse, Universit{\'e} Toulouse 3, INSA-T, Grenoble and Toulouse, France}

\date{\today}

\begin{abstract}
The unique coupling between magnetic order and photo-generated excitons, electron--hole pairs bound by Coulomb interaction, in layered magnetic semiconductors offers a powerful mechanism for controlling light--matter interactions. In the van der Waals antiferromagnet CrSBr, this coupling is exceptionally strong and manifests distinctly between two coexisting excitonic states: the localised, Frenkel-like $X_A$ exciton and the more delocalised, Wannier--Mott-like $X_B$ exciton, providing a unique playground for the optical control of magnetism. Here, we reveal how chlorine incorporation reshapes the magneto-optical interplay in CrSBr$_{1-x}$Cl$_x$ by simultaneously modifying its electronic structure, excitonic properties, and magnetic interactions. Combining magneto-optical spectroscopy up to 85\,T with state-of-the-art quasiparticle self-consistent $GW$ ($\mathrm{QS}G\hat{W}$) calculations on alloy supercells, we show that Cl insertion progressively localises the excitonic wavefunctions and drives both states toward a more Frenkel-like regime. This evolution is accompanied by a systematic reduction of the magnetic-field-induced energy renormalisation, most prominently for the $\mathrm{X_B}$ exciton. Our work connects exciton character directly to magneto-excitonic coupling. Furthermore, it establishes compositional alloying as an effective strategy for engineering the coupling between magnetic and optical properties in van der Waals magnetic semiconductors.
\end{abstract}

\maketitle
\section{Introduction}

The optoelectronic properties of two-dimensional (2D) layered semiconductors are governed by strongly bound excitons, which dominate their optical response and serve as primary probes of their electronic properties\cite{wang2018colloquium,blancon2018scaling,zhang2018determination, goryca2019revealing}. These materials provide a compelling new playground for excitonic physics, hosting a diverse landscape that includes intra- and interlayer excitons\cite{rivera2015observation, hagel2021exciton,leisgang2020giant}, moiré excitons\cite{seyler2019signatures,zhang2018moire, tran2019evidence}, and multiparticle complexes\cite{ross2013electrical, barbone2018charge, van2022six, dijkstra2025ten}, as well as bound electron–hole pairs of fundamentally contrasting character, such as Wannier–Mott and Frenkel-like excitons\cite{smiertka2026distinct,kang2020coherent,grzeszczyk2023strongly,jana2026deconstruction,na2026engineering}. Moreover, the excitonic states in these systems are highly tunable, exhibiting a strong sensitivity not only to external perturbations such as strain\cite{castellanos2013local}, doping \cite{ross2013electrical, barbone2018charge}, and the dielectric environment\cite{raja2017coulomb, raja2019dielectric, stier2016probing}, but also to internal degrees of freedom such as the valley index\cite{xu2014spin} or spontaneous electric polarization\cite{deb2024excitonic, schwandt2025ferroelectric}. Such versatility, alongside strong quasiparticle-mediated light–matter interactions, makes 2D semiconductors ideal testbeds for fundamental physics, enabling the exploration of exotic quantum phases and complex many-body phenomena\cite{chen2022tuning, gu2022dipolar,smolenski2021signatures, huber2026optical, lian2023quadrupolar}.

% \begin{figure*}
%     \centering
%     \includegraphics[width=1\textwidth]{Fig1_draft_v3.png}
%     \caption{\textbf{Optical response of CrSBr$_{1-x}$Cl$_x$.} (a) Low-temperature ($\sim$10\,K) reflectance contrast spectra of CrSBr$_{1-x}$Cl$_x$ flakes measured for different Cl content. The top curve corresponds to pristine CrSBr, and the subsequent curves represent the results for chlorine contents of 20\%, 30\%, 40\%, and 50\% Cl, respectively. The dashed lines mark the positions of two prominent excitonic features labeled \(X_A\) and \(X_B\). (b) Composition dependence of the \(X_A\) and \(X_B\) exciton resonance energies, extracted from reflectance measurements, as a function of Cl content. Dots denote the experimentally determined excitonic feature positions, while diamonds are bandgap values extracted from first principle calculations, and the solid lines are guides to the eye.}
%     \label{fig:Fig1}
% \end{figure*}

The recent emergence of layered magnetic semiconductors introduces new degrees of freedom for manipulating excitonic behaviour by exploiting the intrinsic interplay between magnetic order and optical transitions\cite{Wu19,seyler2018ligand,wilson2021interlayer,dirnberger2023magneto,adak2026excitons}. Such interplay is exceptionally strong in CrSBr, a recent member of the 2D magnetic materials family\cite{goser1990magnetic,ziebel2024crsbr, wilson2021interlayer,henriquez2025strain}. In this van der Waals A-type antiferromagnet, the electronic band structure is highly sensitive to the alignment of the magnetic moments in adjacent layers, which is directly reflected in the optical response\cite{wilson2021interlayer,dirnberger2023magneto,marques2023interplay,bae2022exciton,ruta2023hyperbolic, smiertka2026distinct}. As a result, the low-energy ($X_A$) and high-energy ($X_B$) excitons exhibit energy renormalisations of approximately 10\,meV and 100\,meV, respectively, between the antiferromagnetic (AFM) and ferromagnetic (FM) phases\cite{smiertka2026distinct, shi2024giant, Antoniazzi2026_CrSBr}. The pronounced contrasting shifts of the two transitions have been attributed to the fundamentally different character of the two excitonic states. In particular, CrSBr has been shown to host both: a strongly localised Frenkel-like exciton ($X_A$) and a more delocalized Wannier–Mott-like exciton ($X_B$), making it a rare example of a semiconductor in which excitons with markedly different degrees of localisation coexist \cite{smiertka2026distinct,datta2025magnon}. Such coexistence not only establishes CrSBr as a model system for investigating the interplay between excitonic, magnetic, and lattice degrees of freedom but also raises an intriguing question: can the character of these excitonic states and consequently their coupling to magnetic order be controlled in a systematic manner?

Recently, it has been demonstrated that alloying CrSBr with Cl, forming CrSBr$_{1-x}$Cl$_x$, provides an effective route to tune its magnetic properties \cite{telford2023designing}, including the critical field, N\'eel temperature, and magnetic anisotropy \cite{telford2023designing,badola2026van}. However, the same chemical modifications responsible for this magnetic tuning are also expected to substantially alter the electronic structure of the material and, consequently, its excitonic properties.

Cl substitution alters the CrSBr lattice along two competing chemical axes. On the one hand, the smaller ionic radius of Cl contracts the Cr--halogen bonds, which in isolation would enhance the Cr--halide $p$--$d$ hopping $t$ and broaden the associated bonding/antibonding manifolds. On the other hand, Cl is markedly more electronegative than Br. Replacing Br with Cl shifts the halide p states to lower energies, increasing the charge-transfer energy $\Delta$ between halide-$p$ and Cr-$d$ states. The effective covalency, set by $t^2/\Delta$, is therefore governed by the competition between these two effects, and in CrSBr$_{1-x}$Cl$_x$ the growth of $\Delta$ outweighs the enhancement of $t$, suppressing $p$--$d$ hybridisation and rendering the Cr--ligand network more ionic. This behavious is consistent with the trend already established across the chromium trihalide series CrI$_3$\,$\to$\,CrBr$_3$\,$\to$\,CrCl$_3$, in which increasing halide electronegativity systematically reduces $p$--$d$ hybridisation, narrows the relevant bands, and drives the excitonic states from a more delocalised Wannier-like regime in CrI$_3$ to a strongly localised, atomic-multiplet-like regime in CrCl$_3$~\cite{acharya2021electronic,acharya2022real,grzeszczyk2023strongly}. The result is a tighter, more Cr-localised set of frontier molecular orbitals and a weaker dielectric screening of the on-site electron--hole interaction, both of which push the excitonic states toward the Frenkel limit \cite{kanzaki1971excitons,baranowski2020exciton,grzeszczyk2023strongly}. 

Chemical alloying thus complements other recently demonstrated routes of controlling excitonic localisation in CrSBr, including deterministic atom-by-atom engineering of point defects and defect-bound excitonic states by scanning transmission electron microscopy~\cite{klein2026mesoscale}. Because magnetic order and excitonic properties in CrSBr are inherently coupled, Cl alloying offers a unique opportunity to simultaneously tune both quantities and directly probe their interdependent evolution. Understanding how exciton character, magnetic order, and magneto-excitonic coupling evolve upon chemical substitution is therefore essential for tailoring the optoelectronic and magnetic functionality of this material system.

Here, we elucidate this question through an extensive study of the optical response and its interplay with magnetic order in CrSBr$_{1-x}$Cl$_x$. Combining magneto-optical spectroscopy in extreme magnetic fields up to 85\,T with state-of-the-art electronic structure calculations on alloyed supercells, we track the evolution of both excitonic transitions across the alloy series and reveal how their character changes upon Cl incorporation. We show that increasing Cl content progressively drives the excitonic states toward a more Frenkel-like regime, reducing their spatial extent and modifying their sensitivity to the underlying electronic band structure. This evolution has a direct impact on the magneto-excitonic response of the material. In particular, the high-energy $\mathrm{X_B}$ exciton gradually loses its delocalised Wannier-like character, which results in a reduced energy renormalisation across the AFM--FM phase transition. By correlating the evolution of exciton localisation, magnetic-field response, and electronic structure, we establish a direct link between exciton character and magneto-excitonic coupling. Our results provide a microscopic understanding of the impact of Cl alloying on the coupled magnetic and excitonic properties of CrSBr$_{1-x}$Cl$_x$ and demonstrate that chemical alloying offers a powerful route for engineering magneto-excitonic phenomena in van der Waals semiconductors.

\section{Results}

\begin{figure}
    \centering
    \includegraphics[width=1\columnwidth]{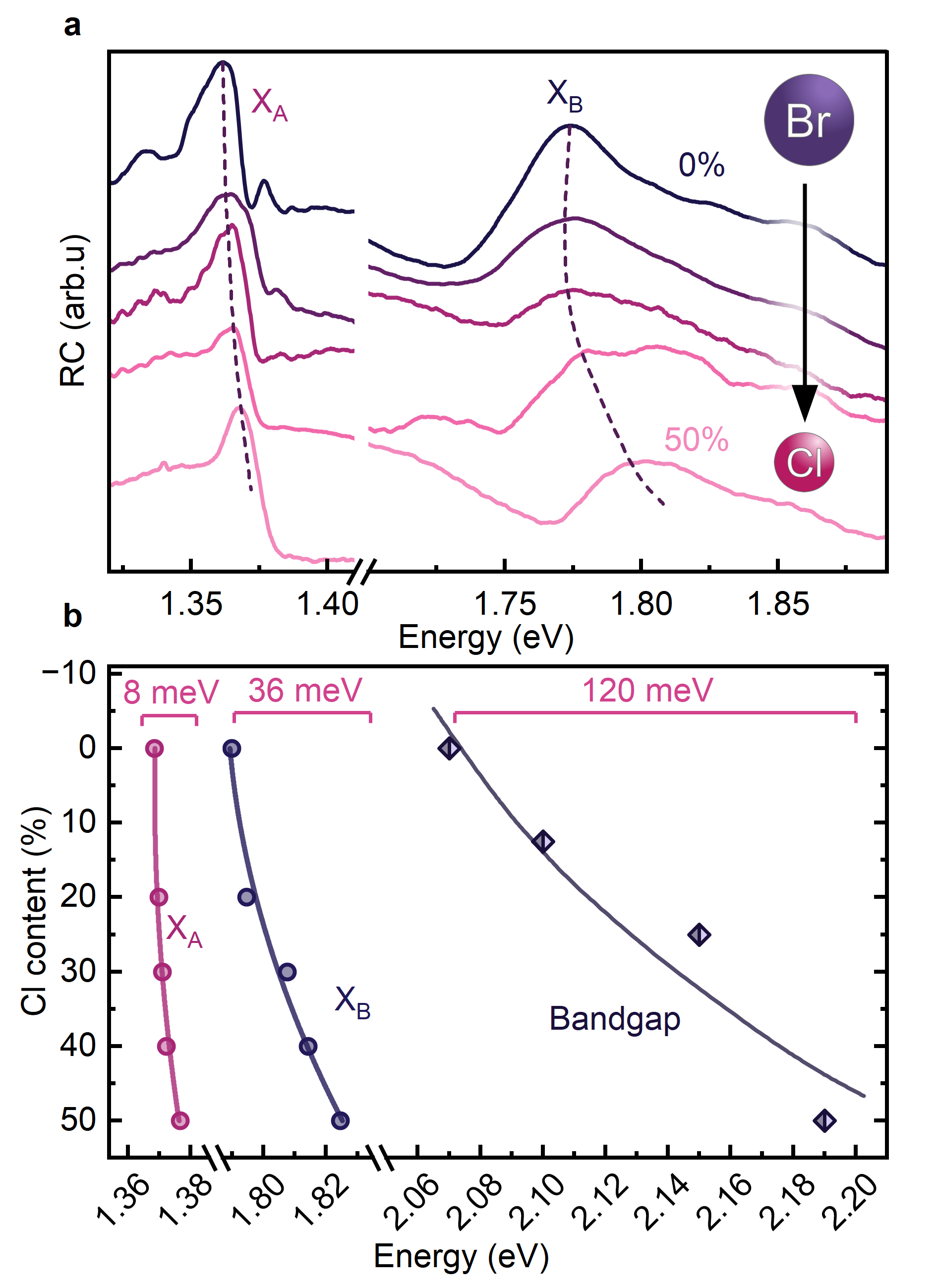}
    \caption{\textbf{Optical response of CrSBr$_{1-x}$Cl$_x$.} (a) Low-temperature ($\sim$10\,K) reflectance contrast spectra of CrSBr$_{1-x}$Cl$_x$ flakes measured for different Cl content. The top curve corresponds to pristine CrSBr, and the subsequent curves represent the results for chlorine contents of 20\%, 30\%, 40\%, and 50\% Cl, respectively. The dashed lines mark the positions of two prominent excitonic features labelled \(X_A\) and \(X_B\). (b) Composition dependence of the \(X_A\) and \(X_B\) exciton resonance energies, extracted from reflectance measurements, as a function of Cl content. Dots denote the experimentally determined excitonic feature positions, while diamonds are bandgap values extracted from first-principles calculations, and the solid lines are guides to the eye.}
    \label{fig:Fig1}
\end{figure}

\begin{figure*}[t]
    \centering
    \includegraphics[width=1\textwidth]{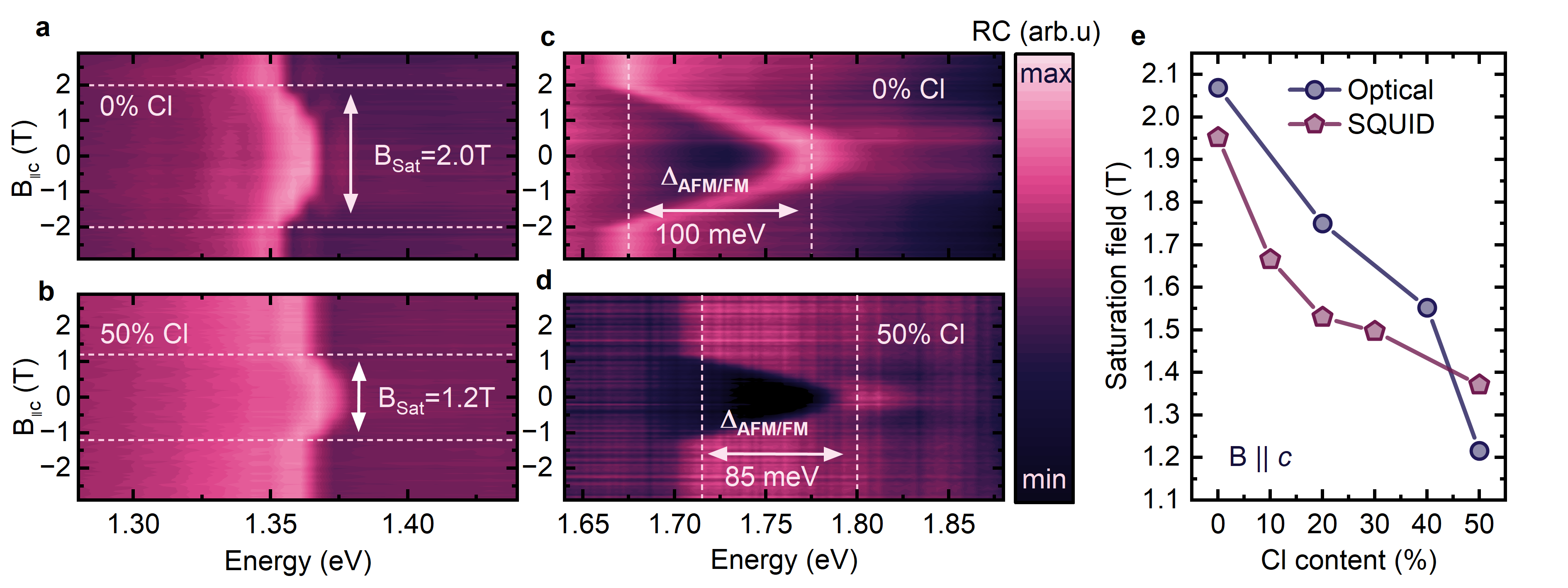}
    \caption{{\textbf{Impact of alloying on the magneto-optical response.} (a-d) Evolution of reflectance contrast spectra around $X_A$ (a,b) and $X_B$ (c,d) energy range as a function of the magnetic field presented in a false-colour map for flakes of pristine CrSBr and CrSBr$_{0.5}$Cl$_{0.5}$. The arrow marks the values of the critical field. (e) Comparison of the magnetic saturation field B\textsubscript{sat} with increasing chlorine concentration, extracted from optical measurements and SQUID. }}
    \label{fig:Fig2}
\end{figure*}

\begin{figure*}[t]
    \centering
    \includegraphics[width=1\textwidth]{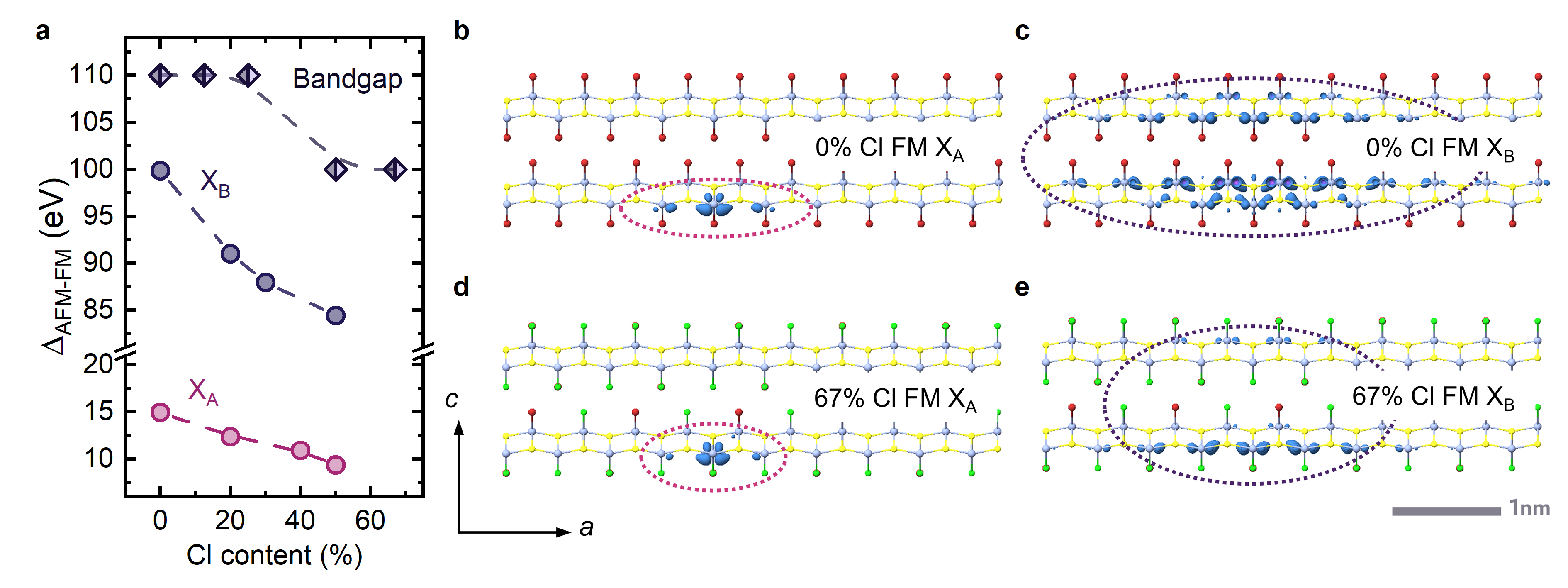}
    \caption{{\textbf{Evolution of AFM/FM energy shifts and simulations of the exciton wavefunction spread with Cl content} (a) Evolution of \(\Delta_{AFM/FM}\) as a function of Cl content for thin flakes. Magenta and purple points represent absolute values of the energy shift between 0 T and 3 T of \(X_A\) and \(X_B\) transitions. Diamonds are the results of the calculated bandgap energy shifts predicted by our QSGW calculations. The dash lines are guides to the eye. (b) Exciton wavefunction isosurface simulations for pristine CrSBr \(X_A\) and \(X_B\) (b-c) and for 67\% Cl (d-e). }}
    \label{fig:Fig3}
\end{figure*}

The results presented in this work are based on an investigation of two types of samples: bulk crystals and exfoliated thin flakes (approximately 10\,nm thick) of CrSBr and CrSBr$_{1-x}$Cl$_x$, with molar chlorine to bromine ratio ranging from 10\% to 50\% (The uniform Cl distribution in the crystals is confirmed by the SEM-EDX characterisation presented in Fig.\,S1). To capture both the magnetic and electronic aspects of the system, we performed magneto-optical spectroscopy across a wide range of magnetic fields, from low fields (0–3\,T) to pulsed high fields reaching up to 85\,T, applied along the hard magnetisation axis (\textit{c}). This approach enables us to connect two complementary regimes: the low-field experiments (performed on exfoliated flakes) provide a detailed view of how the optical response evolves across the magnetic phase transition, while the high-field experiments (performed on bulk crystals) probe the spatial extent of the exciton wavefunction.

The evolution of the optical response with chlorine substitution is shown in Fig.\,\ref{fig:Fig1}a. The reflectance contrast spectrum {(defined as $(R-R_0)/R$, where $R$ is the reflectivity spectrum from the investigated flake and $R_0$ is the reflectivity spectrum from a nearby $\mathrm{SiO_2}$ substrate region)} of CrSBr is dominated by two transitions, $X_A$ and $X_B$, observed at approximately 1.37\,eV and 1.78\,eV, respectively (dark violet curve). Remarkably, both features persist across the entire compositional range, indicating that the fundamental excitonic structure of the material remains robust against chlorine incorporation.

With increasing Cl content, both $X_A$ and $X_B$ exhibit a systematic blueshift, consistent with the calculated increase of the bandgap, as summarised in Fig.~\ref{fig:Fig1}b. While the bandgap transition cannot be reliably resolved in the optical spectra, the evolution of the excitonic resonances qualitatively follows the predicted trend. However, the magnitude of the shift differs substantially between the two excitonic transitions and the predicted bandgap.

Namely, the lower-energy exciton, $X_A$, shifts by only $\sim 8$\,meV even at the highest Cl concentration of 50\%, while the bandgap opens by 120\,meV. This comparatively small change of the $X_A$ energy, despite the pronounced evolution of the bandgap, stems from the more Frenkel-like character of the $X_A$ exciton \cite{smiertka2026distinct,datta2025magnon}. Owing to its strong localisation on the scale of a single atomic site, this transition is only weakly affected by modifications of the host electronic band structure. In contrast, the higher-energy $X_B$ transition exhibits a much stronger response to Cl incorporation, blueshifting by approximately 36\,meV at 50\% Cl content. This behaviour reflects its more Wannier–Mott-like character: given its considerably larger spatial extent in real space \cite{smiertka2026distinct,datta2025magnon}, the $X_B$ exciton is substantially more sensitive to changes in the band-edge energies.

Nevertheless, the blueshift of $X_B$ does not directly track the predicted increase of the bandgap. Even for the highest Cl concentration, the observed shift amounts to only about 30\% of the calculated bandgap change. This discrepancy suggests a progressive increase in the $X_B$ exciton binding energy with increasing Cl content, accompanied by enhanced localisation of its wavefunction. These observations indicate that Cl alloying progressively shifts the character of the $X_B$ exciton toward the Frenkel limit. As discussed in the following sections, this evolution directly impacts the magneto-excitonic coupling in CrSBr$_{1-x}$Cl$_x$.

To investigate the impact of Cl alloying on the magneto-excitonic coupling, we performed magneto-optical spectroscopy measurements. Figure~\ref{fig:Fig2}(a–d) shows the evolution of the excitonic transitions with magnetic field for pristine CrSBr and CrSBr$_{0.5}$Cl$_{0.5}$ (for the remaining Cl composition, see Fig.\,S2). As in pristine CrSBr, in CrSBr$_{0.5}$Cl$_{0.5}$,  both excitons exhibit a characteristic redshift with increasing magnetic field that saturates above a well-defined critical field. This behaviour reflects the field-driven transition from the antiferromagnetic (AFM) to the ferromagnetic (FM) phase, as previously established for CrSBr\cite{wilson2021interlayer,dirnberger2022spin}.

The AFM-to-FM transition manifests itself as a distinct kink in the exciton energy evolution, allowing the critical field to be extracted directly from optical spectroscopy (Fig.~\ref{fig:Fig2}e). The values obtained in this way are in excellent agreement with low-temperature SQUID magnetometry measurements (Fig.~\ref{fig:Fig2}e) performed on large area films of mechanically exfoliated flakes (20-40nm in thickness) \cite{sozen2023high}, confirming that the optical response provides a sensitive probe of the underlying magnetic order. Both techniques reveal a systematic reduction of the critical field with increasing Cl content, from approximately 2\,T in pristine CrSBr to 1.2\,T in CrSBr$_{0.5}$Cl$_{0.5}$. This trend is consistent with the known reduction of magnetocrystalline anisotropy and weakening of interlayer exchange interactions induced by Cl substitution \cite{telford2023designing,badola2026van}.

\begin{figure*}
    \centering
    \includegraphics[scale=1.15]{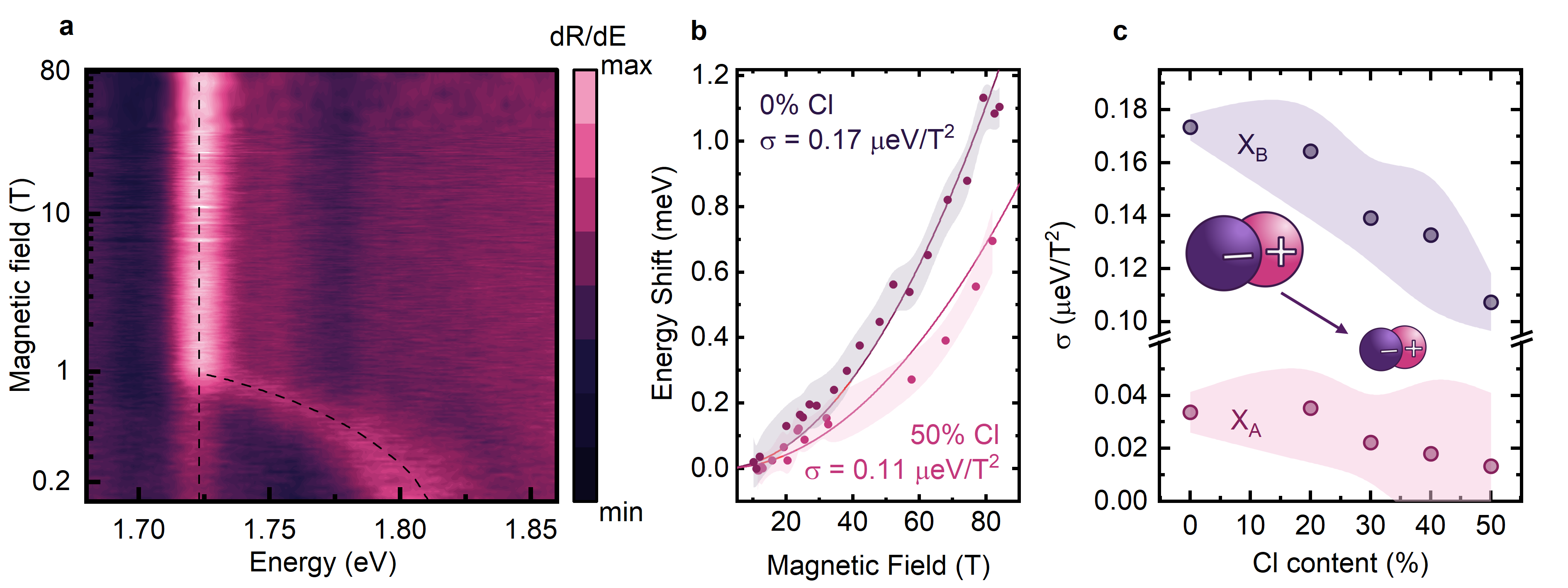}
    \caption{{\textbf{Probing spatial extension of excitonic transitions of CrSBr$_{1-x}$Cl$_x$.} (a) False colour plot of dR/dE and energy vs magnetic field plotted on a logarithmic scale for the \(X_B\) range in high field regime for bulk crystal with 50\% Cl. The dashed lines are guide to the eye (b) Extracted energy shifts in the FM phase for pristine CrSBr and 50\% Cl sample showing a quadratic dependence. The solid lines are parabola fits used to extract the values of the diamagnetic coefficients (c) Diamagnetic coefficients $\sigma$ of $X_A$ and $X_B$ excitons plotted as a function of Cl content. The shading represents an error of the fit to the exciton energy.}}
    \label{fig:Fig4}
\end{figure*}

Crucially, Cl incorporation not only modifies the magnetic and optical properties of CrSBr$_{1-x}$Cl$_x$, but also alters the coupling between them. This effect is directly reflected in the magnetic-field-induced renormalisation of the excitonic transitions. While the high-energy $\mathrm{X_B}$ exciton remains substantially more sensitive to magnetic order than $X_A$, the magnitude of this response decreases systematically with increasing Cl content. In particular, the energy difference between the AFM and FM phases for the $X_B$ transition decreases from $\Delta E_{\mathrm{AFM-FM}} \approx 100$\,meV in pristine CrSBr to approximately 85\,meV for 50\% Cl incorporation (Fig.~\ref{fig:Fig3}a). A similar, albeit weaker, trend is observed for the $X_A$ transition.

To uncover the microscopic origin of this behaviour, we calculated the excitonic wavefunctions in FM phase, as shown in Fig.~\ref{fig:Fig3}b–e (for complementary plots corresponding to the AFM phase, see Fig.\,S3 in SI). Since the structural parameters (bond angles and bond lengths) are adopted from Ref.~\cite{telford2023designing}, the Cl concentrations in the calculations differ slightly from those in the experimentally investigated samples. In both pristine CrSBr and CrSBr$_{0.33}$Cl$_{0.67}$, the $X_A$ exciton remains largely confined within a single layer, consistent with its strongly localised Frenkel-like character. Only a modest increase in localisation is observed upon Cl incorporation. In contrast, the $X_B$ exciton undergoes a much more pronounced transformation. While in pristine CrSBr it exhibits a clear interlayer character (in FM phase), its spatial extent is substantially reduced in CrSBr$_{0.33}$Cl$_{0.67}$, indicating a progressive evolution toward the Frenkel-like limit. This trend provides a natural explanation for the reduced sensitivity of the excitonic transitions to magnetic order. In pristine CrSBr, the exceptionally large energy renormalisation of $X_B$ originates from its delocalised character and strong dependence on the band-edge shifts induced by interlayer hybridisation in the FM phase \cite{smiertka2026distinct}. As Cl alloying drives the excitonic states toward a more localised, Frenkel-like regime, their sensitivity to these band-structure modifications is reduced, resulting in a weaker magneto-excitonic response. The effect is most pronounced for $X_B$, which experiences the largest change in spatial extent, while $X_A$ remains comparatively unaffected owing to its already localised nature.

To experimentally verify the predicted increase in exciton localisation with Cl incorporation, we performed high-field magneto-optical measurements up to 85\,T on the bulk samples. In the ferromagnetic phase, the diamagnetic shift of an excitonic transition provides a direct measure of the spatial extent of the exciton wavefunction in the plane perpendicular to the magnetic field \cite{smiertka2026distinct,PhysRevB.39.10943,stier2016exciton,goryca2019revealing}. As shown in Fig.~\ref{fig:Fig4}a, the rapid energy shift associated with the AFM--FM phase transition at low magnetic fields is followed by a much weaker, yet clearly resolvable, field-induced shift at higher fields.

A detailed analysis of the exciton energy in the high magnetic field regime, presented in Fig.~\ref{fig:Fig4}b for the $X_B$ exciton, reveals the characteristic quadratic dependence expected for the diamagnetic shift of an excitonic transition (results for all samples are provided in the Supporting Information Fig.\,S4):
\begin{equation}
\Delta E=\sigma B^2,
\label{eq:diamagnetic}
\end{equation}
where the diamagnetic coefficient $\sigma$ is related to the in-plane spatial extent of the exciton wavefunction through
\begin{equation}
\sigma=\frac{\mathrm{e}^2}{8\mu}\langle r^2\rangle,
\end{equation}
with $\mu$ denoting the exciton reduced mass and $\langle r^2\rangle$ the expectation value of the squared electron-hole separation perpendicular to the applied magnetic field \cite{goryca2019revealing}.

For comparison, Fig.~\ref{fig:Fig4}b also shows the diamagnetic shift of pristine CrSBr. The substantially smaller diamagnetic coefficient observed in the alloy, and calculated wavefunction isosurfaces in Fig.~\ref{fig:Fig3} strongly suggest a reduced spatial extent of the exciton wavefunction. This trend is systematically observed throughout the CrSBr$_{1-x}$Cl$_x$ series. As summarised in Fig.~\ref{fig:Fig4}c, the diamagnetic coefficients of both $X_A$ and $X_B$ decrease monotonically with increasing Cl content, demonstrating a progressive localisation of the excitonic states. The effect is particularly pronounced for $X_B$, consistent with its evolution from a more Wannier–Mott-like toward a more Frenkel-like character. In contrast, the reduction of the diamagnetic shift is less significant for $X_A$, reflecting its already strong localisation and large binding energy in pristine CrSBr. These observations provide direct experimental confirmation of the wavefunction localisation trend predicted by our electronic structure calculations.

\section{Summary}
In summary, alloying CrSBr with Cl provides a unique platform for tuning the character of excitonic states and, consequently, their coupling to magnetic order. The contrasting response of the $X_A$ and $X_B$ excitons to Cl incorporation highlights their fundamentally different nature. While both transitions blueshift with increasing Cl content, the substantially stronger evolution of $X_B$ reflects its more Wannier--Mott-like character and greater sensitivity to changes in the underlying electronic structure. In contrast, the strongly localised Frenkel-like $X_A$ exciton remains comparatively insensitive to the alloy-induced modification of the band structure.

Our electronic structure calculations, combined with high-field magneto-optical spectroscopy, demonstrate that Cl incorporation progressively localises both excitonic states, driving them toward a more Frenkel-like regime. Microscopically, although the smaller ionic radius of Cl contracts the Cr--halogen bonds and would in isolation enhance Cr--halide $p$--$d$ hopping, the larger electronegativity of Cl deepens the halide $p$ levels and dominates the response: it suppresses $p$--$d$ hybridisation, increases the ionicity of the Cr--ligand network and tightens the local Cr-centred molecular orbitals from which the excitons are built. This localisation is particularly pronounced for $X_B$ and is accompanied by a systematic reduction of its magnetic-field-induced energy renormalisation, establishing a direct link between exciton character and the strength of magneto-excitonic coupling in CrSBr$_{1-x}$Cl$_x$.

More broadly, by continuously tuning the exciton character through chemical alloying, our results extend and experimentally validate the microscopic framework introduced in Ref.~\cite{smiertka2026distinct}. Specifically, we show that the large excitonic energy shifts associated with the AFM--FM phase transition in CrSBr-based materials originate predominantly from magnetic-order-induced modifications of the electronic band structure, to which different excitonic states respond with markedly different sensitivity. Within this picture, the exceptional response of the $X_B$ exciton arises primarily from its more Wannier-like character and its strong dependence on the band-edge energies. This again suggests that $X_B$ should be equally sensitive to other perturbations known to modify the band structure, such as doping, dielectric environment, or strain, making it a particularly rewarding target for future studies in this material family.

Our findings establish chemical alloying as an effective route for engineering coupled magnetic and excitonic properties in layered semiconductors and provide new insight into the interplay between electronic structure, exciton localisation, and magnetic order in van der Waals magnets~\cite{adak2026excitons}.

\section*{Methods}

\subsection*{Sample synthesis}
CrSBr$_{1-x}$Cl$_x$ single crystals with nominal x in the range of 0.1–0.5 were prepared by chemical vapor transport (CVT) in sealed quartz ampoules using chromium (99.99\%, −60 mesh, Chemsavers, USA), bromine (99.9999\%, Merck, Czech Republic), sulfur (99.9999\%, 2–6 mm, Wuhan Tuocai Technology Co. Ltd., China) and chromium(III) chloride (99.9\%, Strem, USA) as starting materials. The chlorine content was targeted up to 50 at.\% on the halogen sublattice. The ampoules were filled with stoichiometric amounts of the starting components corresponding to 16 g of Cl-doped CrSBr, with an additional 2 at.\% excess of sulfur and bromine. The ampoules were sealed under high vacuum while cooling the bromine-containing part with liquid nitrogen.

In the first step, the ampoules were gradually heated to 700 °C in a crucible furnace while keeping the upper part of the ampoule cold to ensure controlled reaction of bromine. After complete consumption of bromine, the ampoules were placed in a two-zone furnace for CVT growth. Initially, the growth zone was heated to 900 °C and the source zone to 750 °C for 2 days. Subsequently, the thermal gradient was reversed, and the source zone was heated to 900 °C while the growth zone was maintained at 800 °C for 14 days. After cooling to room temperature, the ampoules were opened in an argon-filled glovebox.

\subsection*{Optical spectroscopy}
High-field reflectance measurements were performed in a backscattering geometry. The bulk sample was placed in a liquid-helium cryostat inside the coil of a pulsed magnet with a 4\,mm bore diameter. Broadband white light from a tungsten--halogen lamp was guided to the sample through an optical fibre, while the reflected signal was collected by a surrounding fibre bundle and directed to a 500\,mm monochromator equipped with a 300\,gr/mm grating and a back-illuminated EMCCD camera. Spectra were acquired in 1\,ms time frames throughout the 100\,ms pulse. Measurements were repeated for each field range in two spectral windows centred at 900\,nm ($X_A$) and 700\,nm ($X_B$).

Low magnetic field micro-reflectance contrast (RC) measurements were performed in the Faraday configuration, with the magnetic field applied perpendicular to the plane of the sample. The experiments were carried out using a free-beam optical setup integrated with a superconducting magnet generating magnetic fields up to 16\,T. The sample was mounted on an $x$--$y$--$z$ piezoelectric stage and kept at a temperature of 10\,K in a helium gas atmosphere. The spatial resolution of the measurements was on the order of $\sim$1\,$\upmu$m.

Reflectance contrast (RC) measurements were performed using a 100\,W tungsten--halogen lamp as a broadband light source.

\subsection*{Electronic structure and excitonic calculations}
All electronic-structure and excitonic calculations are performed within the Quasiparticle Self-Consistent \textit{GW} approximation~\cite{qsgw,questaal_paper}, $\mathrm{QS}GW$, augmented with ladder diagrams in the screened interaction $W$ ($\mathrm{QS}GW{\rightarrow}\mathrm{QS}G\hat{W}$). Self-consistency removes the starting-point dependence inherent to conventional \textit{GW}, while the inclusion of electron--hole ladders in $W$ enhances the screening, reduces fundamental band gaps and valence bandwidths, and yields consistently high-fidelity band gaps and optical properties for a wide range of systems, including many magnetic insulators~\cite{acharya2021importance,Cunningham2023,acharyaTheoryColorsStrongly2023}. The reliability of this framework for pristine CrSBr was established in our previous work~\cite{smiertka2026distinct}.

For bulk CrSBr in the AFM phase with a 12-atom unit cell we use $a = 3.504$\,\AA, $b = 4.738$\,\AA. Individual layers contain ferromagnetically polarised spins pointing either along the $+b$ or $-b$ axis, while the interlayer coupling is antiferromagnetic. Self-consistency for the single-particle Hamiltonians (LDA, the static quasiparticlised $\mathrm{QS}GW$ and $\mathrm{QS}G\hat{W}$ $\Sigma^0(\mathbf{k})$) is performed on a $10{\times}7{\times}2$ k-mesh, while the relatively smooth dynamical self-energy $\Sigma(\mathbf{k},\omega)$ is constructed using a $6{\times}4{\times}2$ k-mesh. The $\mathrm{QS}GW$ and $\mathrm{QS}G\hat{W}$ cycles are iterated until the RMS change in $\Sigma^0$ reaches $10^{-5}$\,Ry. A two-particle Hamiltonian for the BSE calculation~\cite{Cunningham2023,cunningham2018} of the polarizability, needed for both $\Sigma(\mathbf{k},\omega)$ and the excitonic eigenvalues and eigenfunctions, contains $N_V = 26$ valence and $N_C = 9$ conduction bands. Excitonic eigenvalues of the two-particle Hamiltonian are converged using a $10{\times}7{\times}2$ k-mesh.

The CrSBr$_{1-x}$Cl$_x$ alloys are described by a 96-site supercell of this 12-atom unit cell, comprising 48 atoms (Cr, S, and the halogen sites populated by Br or Cl according to the targeted composition $x$) together with 48 empty spheres, the latter required by the linear muffin-tin orbital basis used in the Questaal implementation~\cite{questaal_paper} to maintain a space-filling representation of the layered, partly open van der Waals structure. The same magnetic configurations as for the pristine 12-atom cell are considered, namely the A-type AFM ground state and the field-induced FM state, treated independently within identical numerical settings. Self-consistency for the single-particle Hamiltonians (LDA, the static quasiparticlised $\mathrm{QS}GW$ and $\mathrm{QS}G\hat{W}$ $\Sigma^0(\mathbf{k})$) is performed on a $10{\times}6{\times}2$ k-mesh, while the relatively smooth dynamical self-energy $\Sigma(\mathbf{k},\omega)$ is constructed using a $5{\times}3{\times}1$ k-mesh. The $\mathrm{QS}GW$ and $\mathrm{QS}G\hat{W}$ cycles are iterated until the RMS change in $\Sigma^0$ reaches $10^{-5}$\,Ry. A two-particle Hamiltonian for the BSE calculation~\cite{Cunningham2023,cunningham2018} of the polarizability, needed for both $\Sigma(\mathbf{k},\omega)$ and the excitonic eigenvalues and eigenfunctions, contains $N_V = 104$ valence and $N_C = 36$ conduction bands, the larger window reflecting the enlarged Brillouin-zone folding of the supercell. Excitonic eigenvalues of the two-particle Hamiltonian are converged using a $10{\times}6{\times}2$ k-mesh. Real-space exciton wavefunction isosurfaces shown in Fig.~\ref{fig:Fig3}b--e are obtained by back-transforming the lowest BSE eigenvectors into the supercell using the same band windows.

\subsection*{SQUID Magnetometry}

SQUID magnetometry measurements were performed on thin-film samples prepared from single crystals of CrSBr$_{1-x}$Cl$_x$, with nominal chlorine substitution levels of (x = 0), 0.10, 0.20, 0.30, and 0.50, corresponding to progressive bromine-to-chlorine substitution. To obtain samples with sufficient material for magnetic characterization while preserving the intrinsic in-plane anisotropy of the parent layered crystals, the films were fabricated by high-throughput mechanical exfoliation using a roll-to-roll setup \cite{sozen2023high,sozen2025wafer}.

Individual CrSBr$_{1-x}$Cl$_x$, crystals were manually positioned on the exfoliation tape with a defined orientation and subjected to repeated unidirectional shearing. This directional exfoliation process promotes cleavage of the crystals into extended thin films while retaining a preferential in-plane crystallographic alignment of the exfoliated flakes \cite{zamora2026roll}. The resulting textured films were transferred onto customized non-magnetic SiO$_2$ substrates, which had been laser-cut to dimensions of 5 mm × 5 mm to ensure comparable sample geometry across the series. For the transfer, the exfoliated tape was laminated onto the SiO$_2$ substrate and heated at 100 °C for 5 min, promoting adhesion of the exfoliated material to the target substrate upon tape removal. To minimize spurious magnetic contributions, all sample preparation steps were carried out without using metallic tools.

Magnetization measurements were performed using a Quantum Design MPMS-5S SQUID magnetometer equipped with a 5 T superconducting magnet. For the measurements reported in this manuscript, the samples were mounted with the substrate surface perpendicular to the applied magnetic field. In this configuration, the magnetic field is applied normal to the exfoliated film plane and perpendicular to the in-plane (a)- and (b)-axes of the aligned flakes.

\subsection*{EDX}
Scanning electron microscopy with energy-dispersive X-ray spectroscopy (SEM-EDX) was performed using a Hitachi SU8000 field-emission scanning electron microscope at an accelerating voltage of 15 kV. Measurements were carried out on thick flakes (>100 nm) to ensure a sufficient EDX signal for reliable compositional analysis.

\section*{Data availability}
The datasets generated and/or analysed during the current study are available from the corresponding authors on reasonable request.

\section*{Acknowledgements}
This work was authored in part by the National Laboratory of the Rockies for the U.S. Department of Energy (DOE) under Contract No.~DE-AC36-08GO28308. For S.A., D.P., and M.v.S., funding was provided by the Computational Chemical Sciences program within the Office of Basic Energy Sciences, U.S. Department of Energy. S.A., D.P., and M.v.S. acknowledge the use of the National Energy Research Scientific Computing Center, under Contract No.~DE-AC02-05CH11231, using NERSC award BES-ERCAP0021783, and also acknowledge that a portion of the research was performed using computational resources sponsored by the Department of Energy's Office of Energy Efficiency and Renewable Energy and located at the National Laboratory of the Rockies, as well as computational resources provided by the Oak Ridge Leadership Computing Facility. The views expressed in the article do not necessarily represent the views of the DOE or the U.S. Government. The U.S. Government retains and the publisher, by accepting the article for publication, acknowledges that the U.S. Government retains a nonexclusive, paid-up, irrevocable, worldwide license to publish or reproduce the published form of this work, or allow others to do so, for U.S. Government purposes. Z.S. and K.M. were supported by project LUAUS25268 from the Ministry of Education, Youth and Sports (MEYS) and by the project Advanced Functional Nanorobots 
ICMM-CSIC authors also acknowledge support from the Severo Ochoa Centres of Excellence program through Grant CEX2024-001445 S, funded by MICIU/AEI/10.13039/501100011033. A.C-G.  acknowledges funding from the European Research Council (ERC) through the ERC-2024 SyG SKIN2DTRONICS project (Grant Agreement No. 101167218). M.G-H. acknowledges funding from the European Research Council (ERC) through the ERC-2024 SyG METRIQS project (Grant Agreement No. 101167432)(reg.~No.~CZ.02.1.01/0.0/0.0/15\_003/0000444 financed by the EFRR).
P.P. and M.\'S. acknowledge the National Science Centre, Poland, within the OPUS programme (Grant No.~2025/57/B/ST3/03197). The publication was also created as part of a project co-financed by the Polish Ministry of Science and Higher Education under contract no.~2025/WK/01 and Polish National Agency for Academic Exchange (agreement no. BPN/BFR/2024/1/00027/U/00001) and Campus France within the Polonium programme.

\section*{Author contributions}

M.Ś. performed high magnetic field measurements, data analysis, drafted the text and figures representing experimental results of the main manuscript and the supplementary information. O.J. performed optical measurements without magnetic field, low magnetic field measurements and contributed to data analysis, manuscript and supplementary information drafting. K.O.-P. and G.Kr. supported low magnetic field measurements. K.P., P.Pe. and A.S. supported high magnetic field measurements. D.P. contributed to the theoretical calculations. K.M. synthesised the CrSBr$_{1-x}$Cl$_x$ crystal with the support and supervision of Z.S. M.v.S. contributed to the theoretical calculation and manuscript writing. Low magnetic field measurements were performed under the supervision of M.R.M. and A.B. G.Ko. performed SEM-EDX characterisation, E.Z-A. performed SQUID magnetometry measurements under the supervision of A.C-G. M.B., together with P.P., conceived the main idea and overall concept of the investigation. They supervised the high-magnetic-field measurements, participated in the analysis and interpretation of the data, and contributed to writing and final shaping of the manuscript. S.A. performed the theoretical calculations and contributed to the manuscript preparation, interpretation of the experimental observations and the development of their microscopic understanding. E.Z-A. fabricated the samples for SQUID magnetometry measurements under the supervision of A.C-G. SQUID magnetometry measurements and analysis was performed by F.M. and M.G-H.

\section*{Competing interests}
The authors declare no competing interests.

\bibliography{Bibliography_Experiment,Bibliography_Theory}

@article{grzeszczyk2023strongly,
  title={Strongly Correlated Exciton-Magnetization System for Optical Spin Pumping in CrBr3 and CrI3.},
  author={Grzeszczyk, M and Acharya, S and Pashov, D and Chen, Z and Vaklinova, K and van Schilfgaarde, M and Watanabe, K and Taniguchi, T and Novoselov, KS and Katsnelson, MI and others},
  journal={Advanced Materials},
  volume={35},
  number={17},
  pages={2209513},
  year={2023},
  publisher={Wiley Online Library}
}

@article{raja2019dielectric,
  title={Dielectric disorder in two-dimensional materials},
  author={Raja, Archana and Waldecker, Lutz and Zipfel, Jonas and Cho, Yeongsu and Brem, Samuel and Ziegler, Jonas D and Kulig, Marvin and Taniguchi, Takashi and Watanabe, Kenji and Malic, Ermin and others},
  journal={Nature nanotechnology},
  volume={14},
  number={9},
  pages={832--837},
  year={2019},
  publisher={Nature Publishing Group UK London}
}

@article{xu2014spin,
  title={Spin and pseudospins in layered transition metal dichalcogenides},
  author={Xu, Xiaodong and Yao, Wang and Xiao, Di and Heinz, Tony F},
  journal={Nature Physics},
  volume={10},
  number={5},
  pages={343--350},
  year={2014},
  publisher={Nature Publishing Group UK London}
}

@article{schwandt2025ferroelectric,
  title={Ferroelectric Control of Interlayer Excitons in 3R-MoS2/MoSe2 Heterostructures},
  author={Schwandt-Krause, Johannes and Miloudi, Mohammed El Amine and Blundo, Elena and Deb, Swarup and Heidkamp, Jan-Niklas and Watanabe, Kenji and Taniguchi, Takashi and Schwartz, Rico and Stier, Andreas and Finley, Jonathan J and others},
  journal={Nano Letters},
  volume={26},
  number={1},
  pages={214--221},
  year={2025},
  publisher={ACS Publications}
}

@article{deb2024excitonic,
  title={Excitonic signatures of ferroelectric order in parallel-stacked MoS2},
  author={Deb, Swarup and Krause, Johannes and Faria Junior, Paulo E and Kempf, Michael Andreas and Schwartz, Rico and Watanabe, Kenji and Taniguchi, Takashi and Fabian, Jaroslav and Korn, Tobias},
  journal={Nature Communications},
  volume={15},
  number={1},
  pages={7595},
  year={2024},
  publisher={Nature Publishing Group UK London}
}

@article{castellanos2013local,
  title={Local strain engineering in atomically thin MoS2},
  author={Castellanos-Gomez, Andres and Rold{\'a}n, Rafael and Cappelluti, Emmanuele and Buscema, Michele and Guinea, Francisco and Van Der Zant, Herre SJ and Steele, Gary A},
  journal={Nano letters},
  volume={13},
  number={11},
  pages={5361--5366},
  year={2013},
  publisher={ACS Publications}
}

@article{stier2016probing,
  title={Probing the influence of dielectric environment on excitons in monolayer WSe2: insight from high magnetic fields},
  author={Stier, Andreas V and Wilson, Nathan P and Clark, Genevieve and Xu, Xiaodong and Crooker, Scott A},
  journal={Nano letters},
  volume={16},
  number={11},
  pages={7054--7060},
  year={2016},
  publisher={ACS Publications}
}

@article{telford2023designing,
  title={Designing magnetic properties in CrSBr through hydrostatic pressure and ligand substitution},
  author={Telford, Evan J and Chica, Daniel G and Ziebel, Michael E and Xie, Kaichen and Manganaro, Nicholas S and Huang, Chun-Ying and Cox, Jordan and Dismukes, Avalon H and Zhu, Xiaoyang and Walsh, James PS and others},
  journal={Advanced Physics Research},
  volume={2},
  number={11},
  pages={2300036},
  year={2023},
  publisher={Wiley Online Library}
}

@article{smiertka2026distinct,
  title={Distinct magneto-optical response of Frenkel and Wannier excitons in CrSBr},
  author={{\'S}miertka, Maciej and Ryga{\l}a, Micha{\l} and Posmyk, Katarzyna and Peksa, Paulina and Dyksik, Mateusz and Pashov, Dimitar and Mosina, Kseniia and Sofer, Zden{\v{e}}k and van Schilfgaarde, Mark and Dirnberger, Florian and others},
  journal={Nature Communications},
  year={2026},
  publisher={Nature Publishing Group UK London}
}

@article{baranowski2020exciton,
  title={Exciton binding energy and effective mass of CsPbCl3: a magneto-optical study},
  author={Baranowski, Michal and Plochocka, Paulina and Su, Rui and Legrand, Laurent and Barisien, Thierry and Bernardot, Frederick and Xiong, Qihua and Testelin, Christophe and Chamarro, Maria},
  journal={Photonics Research},
  volume={8},
  number={10},
  pages={A50--A55},
  year={2020},
  publisher={Chinese Laser Press and Optical Society of America}
}

@article{kanzaki1971excitons,
  title={Excitons in AgBr1-{\ae}Cl{\ae}-transition of relaxed state between free and self-traffed exciton},
  author={Kanzaki, H and Sakuragi, S and Sakamoto, K},
  journal={Solid State Communications},
  volume={9},
  number={13},
  pages={999--1002},
  year={1971},
  publisher={Elsevier}
}

@article{badola2026van,
  title={Van der Waals CrSBr alloys with tunable magnetic and optical properties},
  author={Badola, Shalini and Pawbake, Amit and Wu, Bing and Soll, Aljoscha and Sofer, Zdenek and Heid, Rolf and Faugeras, Clement},
  journal={Nano Letters},
  year={2026},
  publisher={ACS Publications}
}

@article{goryca2019revealing,
  title={Revealing exciton masses and dielectric properties of monolayer semiconductors with high magnetic fields},
  author={Goryca, Mateusz and Li, Jing and Stier, Andreas V and Taniguchi, Takashi and Watanabe, Kenji and Courtade, Emmanuel and Shree, Shivangi and Robert, Cedric and Urbaszek, Bernhard and Marie, Xavier and others},
  journal={Nature communications},
  volume={10},
  number={1},
  pages={4172},
  year={2019},
  publisher={Nature Publishing Group UK London}
}

@article{zhang2018moire,
  title={Moir{\'e} intralayer excitons in a MoSe2/MoS2 heterostructure},
  author={Zhang, Nan and Surrente, Alessandro and Baranowski, Micha{\l} and Maude, Duncan K and Gant, Patricia and Castellanos-Gomez, Andres and Plochocka, Paulina},
  journal={Nano letters},
  volume={18},
  number={12},
  pages={7651--7657},
  year={2018},
  publisher={ACS Publications}
}

@article{seyler2019signatures,
  title={Signatures of moir{\'e}-trapped valley excitons in MoSe2/WSe2 heterobilayers},
  author={Seyler, Kyle L and Rivera, Pasqual and Yu, Hongyi and Wilson, Nathan P and Ray, Essance L and Mandrus, David G and Yan, Jiaqiang and Yao, Wang and Xu, Xiaodong},
  journal={Nature},
  volume={567},
  number={7746},
  pages={66--70},
  year={2019},
  publisher={Nature Publishing Group UK London}
}

@article{rivera2015observation,
  title={Observation of long-lived interlayer excitons in monolayer MoSe2--WSe2 heterostructures},
  author={Rivera, Pasqual and Schaibley, John R and Jones, Aaron M and Ross, Jason S and Wu, Sanfeng and Aivazian, Grant and Klement, Philip and Seyler, Kyle and Clark, Genevieve and Ghimire, Nirmal J and others},
  journal={Nature communications},
  volume={6},
  number={1},
  pages={6242},
  year={2015},
  publisher={Nature Publishing Group UK London}
}

@article{leisgang2020giant,
  title={Giant Stark splitting of an exciton in bilayer MoS2},
  author={Leisgang, Nadine and Shree, Shivangi and Paradisanos, Ioannis and Sponfeldner, Lukas and Robert, Cedric and Lagarde, Delphine and Balocchi, Andrea and Watanabe, Kenji and Taniguchi, Takashi and Marie, Xavier and others},
  journal={Nature nanotechnology},
  volume={15},
  number={11},
  pages={901--907},
  year={2020},
  publisher={Nature Publishing Group UK London}
}

@article{hagel2021exciton,
  title={Exciton landscape in van der Waals heterostructures},
  author={Hagel, Joakim and Brem, Samuel and Linder{\"a}lv, Christopher and Erhart, Paul and Malic, Ermin},
  journal={Physical Review Research},
  volume={3},
  number={4},
  pages={043217},
  year={2021},
  publisher={APS}
}

@article{zhang2018determination,
  title={Determination of layer-dependent exciton binding energies in few-layer black phosphorus},
  author={Zhang, Guowei and Chaves, Andrey and Huang, Shenyang and Wang, Fanjie and Xing, Qiaoxia and Low, Tony and Yan, Hugen},
  journal={Science advances},
  volume={4},
  number={3},
  pages={eaap9977},
  year={2018},
  publisher={American Association for the Advancement of Science}
}

@article{stier2016exciton,
  title={Exciton diamagnetic shifts and valley Zeeman effects in monolayer WS2 and MoS2 to 65 Tesla},
  author={Stier, Andreas V and McCreary, Kathleen M and Jonker, Berend T and Kono, Junichiro and Crooker, Scott A},
  journal={Nature communications},
  volume={7},
  number={1},
  pages={10643},
  year={2016},
  publisher={Nature Publishing Group UK London}
}

@article{van2022six,
  title={Six-body and eight-body exciton states in monolayer WSe 2},
  author={Van Tuan, Dinh and Shi, Su-Fei and Xu, Xiaodong and Crooker, Scott A and Dery, Hanan},
  journal={Physical Review Letters},
  volume={129},
  number={7},
  pages={076801},
  year={2022},
  publisher={APS}
}

@article{barbone2018charge,
  title={Charge-tuneable biexciton complexes in monolayer WSe2},
  author={Barbone, Matteo and Montblanch, Alejandro R-P and Kara, Dhiren M and Palacios-Berraquero, Carmen and Cadore, Alisson R and De Fazio, Domenico and Pingault, Benjamin and Mostaani, Elaheh and Li, Han and Chen, Bin and others},
  journal={Nature communications},
  volume={9},
  number={1},
  pages={3721},
  year={2018},
  publisher={Nature Publishing Group UK London}
}

@article{dijkstra2025ten,
  title={Ten-valley excitonic complexes in charge-tunable monolayer WSe2},
  author={Dijkstra, Alain and Ben Mhenni, Amine and Van Tuan, Dinh and {\c{C}}etiner, Elif and Schur-Wilkens, Muriel and Kim, Junghwan and Steiner, Laurin and Watanabe, Kenji and Taniguchi, Takashi and Barbone, Matteo and others},
  journal={Nature Communications},
  volume={16},
  number={1},
  pages={9743},
  year={2025},
  publisher={Nature Publishing Group UK London}
}

@article{ross2013electrical,
  title={Electrical control of neutral and charged excitons in a monolayer semiconductor},
  author={Ross, Jason S and Wu, Sanfeng and Yu, Hongyi and Ghimire, Nirmal J and Jones, Aaron M and Aivazian, Grant and Yan, Jiaqiang and Mandrus, David G and Xiao, Di and Yao, Wang and others},
  journal={Nature communications},
  volume={4},
  number={1},
  pages={1474},
  year={2013},
  publisher={Nature Publishing Group UK London}
}

@article{datta2025magnon,
  title={Magnon-mediated exciton--exciton interaction in a van der Waals antiferromagnet},
  author={Datta, Biswajit and Adak, Pratap Chandra and Yu, Sichao and Valiyaparambil Dharmapalan, Agneya and Hall, Siedah J and Vakulenko, Anton and Komissarenko, Filipp and Kurganov, Egor and Quan, Jiamin and Wang, Wei and others},
  journal={Nature Materials},
  pages={1--7},
  year={2025},
  publisher={Nature Publishing Group UK London}
}

@article{shi2024giant,
  title={Giant magneto-exciton coupling in 2D van der Waals CrSBr},
  author={Shi, Jia and Wang, Dan and Jiang, Nai and Xin, Ziqian and Zheng, Houzhi and Shen, Chao and Zhang, Xinping and Liu, Xinfeng},
  journal={ACS nano},
  year={2024},
  publisher={ACS Publications}
}

@article{dirnberger2023magneto,
  title={Magneto-optics in a van der Waals magnet tuned by self-hybridized polaritons},
  author={Dirnberger, Florian and Quan, Jiamin and Bushati, Rezlind and Diederich, Geoffrey M and Florian, Matthias and Klein, Julian and Mosina, Kseniia and Sofer, Zdenek and Xu, Xiaodong and Kamra, Akashdeep and others},
  journal={Nature},
  volume={620},
  number={7974},
  pages={533--537},
  year={2023},
  publisher={Nature Publishing Group UK London}
}

@article{wilson2021interlayer,
  title={Interlayer electronic coupling on demand in a 2D magnetic semiconductor},
  author={Wilson, Nathan P and Lee, Kihong and Cenker, John and Xie, Kaichen and Dismukes, Avalon H and Telford, Evan J and Fonseca, Jordan and Sivakumar, Shivesh and Dean, Cory and Cao, Ting and others},
  journal={Nature Materials},
  volume={20},
  number={12},
  pages={1657--1662},
  year={2021},
  publisher={Nature Publishing Group UK London}
}

@article{seyler2018ligand,
  title={Ligand-field helical luminescence in a 2D ferromagnetic insulator},
  author={Seyler, Kyle L and Zhong, Ding and Klein, Dahlia R and Gao, Shiyuan and Zhang, Xiaoou and Huang, Bevin and Navarro-Moratalla, Efr{\'e}n and Yang, Li and Cobden, David H and McGuire, Michael A and others},
  journal={Nature Physics},
  volume={14},
  number={3},
  pages={277--281},
  year={2018},
  publisher={Nature Publishing Group}
}

@article{bae2022exciton,
  title={Exciton-coupled coherent magnons in a 2D semiconductor},
  author={Bae, Youn Jue and Wang, Jue and Scheie, Allen and Xu, Junwen and Chica, Daniel G and Diederich, Geoffrey M and Cenker, John and Ziebel, Michael E and Bai, Yusong and Ren, Haowen and others},
  journal={Nature},
  volume={609},
  number={7926},
  pages={282--286},
  year={2022},
  publisher={Nature Publishing Group UK London}
}

@article{ziebel2024crsbr,
  title={CrSBr: an air-stable, two-dimensional magnetic semiconductor},
  author={Ziebel, Michael E and Feuer, Margalit L and Cox, Jordan and Zhu, Xiaoyang and Dean, Cory R and Roy, Xavier},
  journal={Nano Letters},
  volume={24},
  number={15},
  pages={4319--4329},
  year={2024},
  publisher={ACS Publications}
}

@article{dirnberger2022spin,
  title={Spin-correlated exciton--polaritons in a van der Waals magnet},
  author={Dirnberger, Florian and Bushati, Rezlind and Datta, Biswajit and Kumar, Ajesh and MacDonald, Allan H and Baldini, Edoardo and Menon, Vinod M},
  journal={Nature Nanotechnology},
  volume={17},
  number={10},
  pages={1060--1064},
  year={2022},
  publisher={Nature Publishing Group UK London}
}

@article{kang2020coherent,
  title={Coherent many-body exciton in van der Waals antiferromagnet NiPS3},
  author={Kang, Soonmin and Kim, Kangwon and Kim, Beom Hyun and Kim, Jonghyeon and Sim, Kyung Ik and Lee, Jae-Ung and Lee, Sungmin and Park, Kisoo and Yun, Seokhwan and Kim, Taehun and others},
  journal={Nature},
  volume={583},
  number={7818},
  pages={785--789},
  year={2020},
  publisher={Nature Publishing Group UK London}
}

@article{tran2019evidence,
  title={Evidence for moir{\'e} excitons in van der Waals heterostructures},
  author={Tran, Kha and Moody, Galan and Wu, Fengcheng and Lu, Xiaobo and Choi, Junho and Kim, Kyounghwan and Rai, Amritesh and Sanchez, Daniel A and Quan, Jiamin and Singh, Akshay and others},
  journal={Nature},
  volume={567},
  number={7746},
  pages={71--75},
  year={2019},
  publisher={Nature Publishing Group UK London}
}

@article{raja2017coulomb,
  title={Coulomb engineering of the bandgap and excitons in two-dimensional materials},
  author={Raja, Archana and Chaves, Andrey and Yu, Jaeeun and Arefe, Ghidewon and Hill, Heather M and Rigosi, Albert F and Berkelbach, Timothy C and Nagler, Philipp and Sch{\"u}ller, Christian and Korn, Tobias and others},
  journal={Nature communications},
  volume={8},
  number={1},
  pages={15251},
  year={2017},
  publisher={Nature Publishing Group UK London}
}

@article{smolenski2021signatures,
  title={Signatures of Wigner crystal of electrons in a monolayer semiconductor},
  author={Smole{\'n}ski, Tomasz and Dolgirev, Pavel E and Kuhlenkamp, Clemens and Popert, Alexander and Shimazaki, Yuya and Back, Patrick and Lu, Xiaobo and Kroner, Martin and Watanabe, Kenji and Taniguchi, Takashi and others},
  journal={Nature},
  volume={595},
  number={7865},
  pages={53--57},
  year={2021},
  publisher={Nature Publishing Group UK London}
}

@article{lian2023quadrupolar,
  title={Quadrupolar excitons and hybridized interlayer Mott insulator in a trilayer moir{\'e} superlattice},
  author={Lian, Zhen and Chen, Dongxue and Ma, Lei and Meng, Yuze and Su, Ying and Yan, Li and Huang, Xiong and Wu, Qiran and Chen, Xinyue and Blei, Mark and others},
  journal={Nature communications},
  volume={14},
  number={1},
  pages={4604},
  year={2023},
  publisher={Nature Publishing Group UK London}
}

@article{huber2026optical,
  title={Optical control over topological Chern number in moir{\'e} materials},
  author={Huber, Olivier and Kuhlbrodt, Kilian and Anderson, Eric and Li, Weijie and Watanabe, Kenji and Taniguchi, Takashi and Kroner, Martin and Xu, Xiaodong and Imamo{\u{g}}lu, A and Smole{\'n}ski, T},
  journal={Nature},
  volume={649},
  number={8099},
  pages={1153--1158},
  year={2026},
  publisher={Nature Publishing Group UK London}
}

@article{gu2022dipolar,
  title={Dipolar excitonic insulator in a moir{\'e} lattice},
  author={Gu, Jie and Ma, Liguo and Liu, Song and Watanabe, Kenji and Taniguchi, Takashi and Hone, James C and Shan, Jie and Mak, Kin Fai},
  journal={Nature physics},
  volume={18},
  number={4},
  pages={395--400},
  year={2022},
  publisher={Nature Publishing Group UK London}
}

@article{chen2022tuning,
  title={Tuning moir{\'e} excitons and correlated electronic states through layer degree of freedom},
  author={Chen, Dongxue and Lian, Zhen and Huang, Xiong and Su, Ying and Rashetnia, Mina and Yan, Li and Blei, Mark and Taniguchi, Takashi and Watanabe, Kenji and Tongay, Sefaattin and others},
  journal={Nature communications},
  volume={13},
  number={1},
  pages={4810},
  year={2022},
  publisher={Nature Publishing Group UK London}
}

@article{blancon2018scaling,
  title={Scaling law for excitons in 2D perovskite quantum wells},
  author={Blancon, J-C and Stier, Andreas V and Tsai, Hsinhan and Nie, Wanyi and Stoumpos, Costas C and Traore, Boubacar and Pedesseau, L and Kepenekian, Mikael and Katsutani, Fumiya and Noe, GT and others},
  journal={Nature communications},
  volume={9},
  number={1},
  pages={2254},
  year={2018},
  publisher={Nature Publishing Group UK London}
}

@article{wang2018colloquium,
  title={Colloquium: Excitons in atomically thin transition metal dichalcogenides},
  author={Wang, Gang and Chernikov, Alexey and Glazov, Mikhail M and Heinz, Tony F and Marie, Xavier and Amand, Thierry and Urbaszek, Bernhard},
  journal={Reviews of Modern Physics},
  volume={90},
  number={2},
  pages={021001},
  year={2018},
  publisher={APS}
}

@article{marques2023interplay,
  title={Interplay between optical emission and magnetism in the van der Waals magnetic semiconductor CrSBr in the two-dimensional limit},
  author={Marques-Moros, Francisco and Boix-Constant, Carla and Ma{\~n}as-Valero, Samuel and Canet-Ferrer, Josep and Coronado, Eugenio},
  journal={ACS nano},
  volume={17},
  number={14},
  pages={13224--13231},
  year={2023},
  publisher={ACS Publications}
}

@article{goser1990magnetic,
  title={Magnetic properties of CrSBr},
  author={G{\"o}ser, O and Paul, W and Kahle, HG},
  journal={Journal of magnetism and magnetic materials},
  volume={92},
  number={1},
  pages={129--136},
  year={1990},
  publisher={Elsevier}
}

@article{Antoniazzi2026_CrSBr,
  author = {Antoniazzi, I. and Kipczak, {\L}ucja and Camargo, B. and Gayatri and Mohanty, C. and Ali, W. and Mosina, K. and Sofer, Z. and Babiński, A. and Karmakar, A. and Molas, M. R.},
  title = {Magneto-excitonic duality from monolayer to trilayer CrSBr},
  journal = {2D Materials},
  year = {2026},
  volume = {13},
  number = {1},
  pages = {015032},
  doi = {10.1088/2053-1583/ae34c1}
}

@article{henriquez2025strain,
  title={Strain Engineering of Magnetoresistance and Magnetic Anisotropy in CrSBr},
  author={Henr{\'\i}quez-Guerra, Eudomar and Ruiz, Alberto M and Galbiati, Marta and Cort{\'e}s-Flores, {\'A}lvaro and Brown, Daniel and Zamora-Amo, Esteban and Almonte, Lisa and Shumilin, Andrei and Salvador-S{\'a}nchez, Juan and P{\'e}rez-Rodr{\'\i}guez, Ana and others},
  journal={Advanced Materials},
  pages={2506695},
  year={2025},
  publisher={Wiley Online Library}
}

@article{sozen2023high,
  title={High-Throughput Mechanical Exfoliation for Low-Cost Production of van der Waals Nanosheets},
  author={Sozen, Yigit and Riquelme, Juan J and Xie, Yong and Munuera, Carmen and Castellanos-Gomez, Andres},
  journal={Small Methods},
  volume={7},
  number={10},
  pages={2300326},
  year={2023},
  publisher={Wiley Online Library}
}

@article{sozen2025wafer,
  title={Wafer-Scale Films of Two-Dimensional Materials via Roll-to-Roll Mechanical Exfoliation},
  author={Sozen, Yigit and Pucher, Thomas and Kesavan, Bhagyanath Paliyottil and Jimenez-Arevalo, Nuria and Hernandez-Ruiz, Julia and Sofer, Zdenek and Munuera, Carmen and Riquelme, Juan J and Castellanos-Gomez, Andres},
  journal={arXiv preprint arXiv:2511.06960},
  year={2025}
}

@article{zamora2026roll,
  title={Roll-to-Roll Mechanical Exfoliation for Large-Area van der Waals Films with Preserved Crystallographic Alignment},
  author={Zamora-Amo, Esteban and Fernandez-Canizares, Francisco and Sozen, Yigit and Pucher, Thomas and Ruano, Pedro L Alc{\'a}zar and Munuera, Carmen and Ishikawa, Ryo and Shibata, Naoya and Varela, Mar{\'\i}a and Lin, Der-Yuh and others},
  journal={Advanced Functional Materials},
  volume={36},
  number={25},
  pages={e19080},
  year={2026},
  publisher={Wiley Online Library}
}

@article{qsgw,  title={Quasiparticle self-consistent g w theory},
  author={van Schilfgaarde, Mark and Kotani, Takao and Faleev, Sergey},
  journal={Physical review letters},
  volume={96},
  number={22},
  pages={226402},
  year={2006},
  url={https://doi.org/10.1103/PhysRevLett.96.226402},
  publisher={APS}
}

@article{PhysRevB.39.10943,
  title = {Diamagnetism as a probe of exciton localization in quantum wells},
  author = {Nash, K. J. and Skolnick, M. S. and Claxton, P. A. and Roberts, J. S.},
  journal = {Physical Review B},
  volume = {39},
  issue = {15},
  pages = {10943--10954},
  numpages = {0},
  year = {1989},
  month = {May},
  publisher = {American Physical Society},
  doi = {10.1103/PhysRevB.39.10943},
  url = {https://link.aps.org/doi/10.1103/PhysRevB.39.10943}
}

@article{cunningham2018,
  title={Effect of ladder diagrams on optical absorption spectra in a quasiparticle self-consistent GW framework},
  author={Cunningham, Brian and Gr{\"u}ning, Myrta and Azarhoosh, Pooya and Pashov, Dimitar and Van Schilfgaarde, Mark},
  journal={Physical Review Materials},
  volume={2},
  number={3},
  pages={034603},
  year={2018},
  publisher={APS}
,
  doi     = {10.1103/PhysRevMaterials.2.034603},
  url     = {https://doi.org/10.1103/PhysRevMaterials.2.034603}
}

@article{Cunningham2023,
   author = {Brian Cunningham and Myrta Gr{\"u}ning and Dimitar Pashov and Mark van Schilfgaarde},
   year = "2023",
   journal = "Phys. Rev. B",
   title = {{QS$G\hat{W}$: Quasiparticle Self Consistent $GW$ with Ladder Diagrams in $W$}},
   volume = "108",
   pages = "165104",
   url = {https://journals.aps.org/prb/abstract/10.1103/PhysRevB.108.165104}
}

@ARTICLE{questaal_paper,
   author = "Dimitar Pashov and Swagata Acharya and Walter R. L. Lambrecht and Jerome Jackson and Kirill D. Belashchenko and Athanasios Chantis and Francois Jamet and Mark van Schilfgaarde",
   year = "2020",
   journal = "Comp. Phys. Comm.",
   title = {{Questaal: a package of electronic structure methods based on the linear muffin-tin orbital technique}},
   volume = "249",
   pages = "107065",
   anote = {},
   url={https://www.sciencedirect.com/science/article/pii/S0010465519303868?via%3Dihub}
}

@article{acharya2021importance,
  title={Importance of charge self-consistency in first-principles description of strongly correlated systems},
  author={Acharya, Swagata and Pashov, Dimitar and Rudenko, Alexander N and R{\"o}sner, Malte and van Schilfgaarde, Mark and Katsnelson, Mikhail I},
  journal={npj Computational Materials},
  volume={7},
  number={1},
  pages={1--8},
  year={2021},
  url={https://doi.org/10.1038/s41524-021-00676-5},
  publisher={Nature Publishing Group}
}

@article{acharya2021electronic,  title={Electronic structure of chromium trihalides beyond density functional theory},
  author={Acharya, Swagata and Pashov, Dimitar and Cunningham, Brian and Rudenko, Alexander N and R{\"o}sner, Malte and Gr{\"u}ning, Myrta and van Schilfgaarde, Mark and Katsnelson, Mikhail I},
  journal={Physical Review B},
  volume={104},
  number={15},
  pages={155109},
  year={2021},
  doi={10.1103/PhysRevB.104.155109},
  url={https://doi.org/10.1103/PhysRevB.104.155109},
  publisher={APS}
}

@article{acharya2022real,  title={Real-and momentum-space description of the excitons in bulk and monolayer chromium tri-halides},
  author={Acharya, Swagata and Pashov, Dimitar and Rudenko, Alexander N and R{\"o}sner, Malte and Schilfgaarde, Mark van and Katsnelson, Mikhail I},
  journal={npj 2D Materials and Applications},
  volume={6},
  number={1},
  pages={1--10},
  year={2022},
  doi={10.1038/s41699-022-00307-7},
  url={https://doi.org/10.1038/s41699-022-00307-7},
  publisher={Nature Publishing Group}
}

@article{acharyaTheoryColorsStrongly2023,
  title = {A Theory for Colors of Strongly Correlated Electronic Systems},
  author = {Acharya, Swagata and Pashov, Dimitar and Weber, Cedric and {van Schilfgaarde}, Mark and Lichtenstein, Alexander I. and Katsnelson, Mikhail I.},
  year = {2023},
  month = sep,
  journal = {Nature Communications},
  volume = {14},
  number = {1},
  pages = {5565},
  publisher = {Nature Publishing Group},
  issn = {2041-1723},
  doi = {10.1038/s41467-023-41314-6},
  urldate = {2024-12-03},
  copyright = {2023 The Author(s)},
  langid = {english},
  url={https://doi.org/10.1038/s41467-023-41314-6}
}

@article{ruta2023hyperbolic,
  title={Hyperbolic exciton polaritons in a van der Waals magnet},
  author={Ruta, Francesco L and Zhang, Shuai and Shao, Yinming and Moore, Samuel L and Acharya, Swagata and Sun, Zhiyuan and Qiu, Siyuan and Geurs, Johannes and Kim, Brian SY and Fu, Matthew and others},
  journal={Nature Communications},
  volume={14},
  number={1},
  pages={8261},
  year={2023},
  url={https://doi.org/10.1038/s41467-023-44100-6},
  publisher={Nature Publishing Group UK London}
}

@ARTICLE{Wu19,
   author = {M. Wu and Z. Li and T. Cao and S. G. Louie},
   year = "2019",
   journal = "Nat Commun",
   title = {{Physical origin of giant excitonic and magneto-optical responses in two-dimensional ferromagnetic insulators.}},
   volume = "10",
   pages = "2371",
   doi = {https://doi.org/10.1038/s41467-019-10325-7},
   url = {https://doi.org/10.1038/s41467-019-10325-7}
}

@article{klein2026mesoscale,
  title   = {Mesoscale atomic engineering in a crystal lattice},
  author  = {Klein, Julian and Roccapriore, Kevin M. and Weile, Mads and Grytsiuk, Sergii and Lupini, Andrew R. and Sofer, Zdenek and Pashov, Dimitar and van Schilfgaarde, Mark and Acharya, Swagata and R{\"o}sner, Malte and Ross, Frances M.},
  journal = {Nature},
  volume  = {653},
  number  = {8115},
  pages   = {715--722},
  year    = {2026},
  doi     = {10.1038/s41586-026-10431-9},
  url     = {https://doi.org/10.1038/s41586-026-10431-9}
}

@article{adak2026excitons,
  title         = {Excitons in van der {W}aals magnetic materials},
  author        = {Adak, Pratap Chandra and Dirnberger, Florian and Acharya, Swagata and Kamra, Akashdeep and Xu, Xiaodong and Menon, Vinod M.},
  journal       = {Nature Materials},
  year          = {2026},
  note          = {in press},
  eprint        = {2602.10409},
  archivePrefix = {arXiv},
  primaryClass  = {cond-mat.mtrl-sci},
  url           = {https://arxiv.org/abs/2602.10409}
}

@article{na2026engineering,
  title         = {Engineering photomagnetism in collinear van der {W}aals antiferromagnets},
  author        = {Na, MengXing and Radovskaia, Viktoriia and Khusyainov, Dinar and Kim, Peter and Mukhuti, Kingshuk and Christianen, Peter C. M. and Kochetkova, Ekaterina and Isaeva, Anna and de Visser, Anne and Pashov, Dimitar and van Schilfgaarde, Mark and Teo, Edwin H. T. and Chaturvedi, Apoorva and Acharya, Swagata and Rasing, Theo and Kimel, Alexey V. and Afanasiev, Dmytro},
  journal       = {arXiv preprint arXiv:2603.10186},
  year          = {2026},
  eprint        = {2603.10186},
  archivePrefix = {arXiv},
  primaryClass  = {cond-mat.mtrl-sci},
  url           = {https://arxiv.org/abs/2603.10186}
}

@article{jana2026deconstruction,
  title   = {Deconstruction of the Anisotropic Magnetic Interactions from Spin-Entangled Optical Excitations in van der {W}aals Antiferromagnets},
  author  = {Jana, Dipankar and Acharya, Swagata and Orlita, Milan and Faugeras, Cl{\'e}ment and Pashov, Dimitar and van Schilfgaarde, Mark and Potemski, Marek and Koperski, Maciej},
  journal = {Advanced Science},
  volume  = {13},
  number  = {2},
  pages   = {e05834},
  year    = {2026},
  doi     = {10.1002/advs.202505834},
  url     = {https://doi.org/10.1002/advs.202505834},
  publisher = {Wiley}
}

\end{document}

\title{Tunable Magneto-Excitonic Coupling in Alloyed van der Waals Antiferromagnet}

%Tuning the Wannier-Mott to Frenkel character of excitons through halogen substitution in CrSBr
%Chemical tuning of the Wannier-Mott to Frenkel character of excitons in CrSBr

\author{Maciej Śmiertka}
\affiliation{Department of Experimental Physics, Faculty of Fundamental Problems of Technology, Wroclaw University of Science and Technology, 50-370 Wroclaw, Poland}

\author{Oliwia Janikowska}
\affiliation{Department of Experimental Physics, Faculty of Fundamental Problems of Technology, Wroclaw University of Science and Technology, 50-370 Wroclaw, Poland}

\author{Katarzyna Olkowska-Pucko}
\affiliation{Faculty of Physics, University of Warsaw, 02-093 Warsaw, Poland}

\author{Grzegorz Krasucki}
\affiliation{Faculty of Physics, University of Warsaw, 02-093 Warsaw, Poland}

\author{Katarzyna Posmyk}
\affiliation{Department of Experimental Physics, Faculty of Fundamental Problems of Technology, Wroclaw University of Science and Technology, 50-370 Wroclaw, Poland}
\affiliation{Laboratoire National des Champs Magn\'etiques Intenses, EMFL, CNRS UPR 3228, Universit{\'e} Grenoble Alpes, Universit{\'e} Toulouse, Universit{\'e} Toulouse 3, INSA-T, Grenoble and Toulouse, France}

\author{Paulina Peksa}
\affiliation{Department of Experimental Physics, Faculty of Fundamental Problems of Technology, Wroclaw University of Science and Technology, 50-370 Wroclaw, Poland}
\affiliation{Laboratoire National des Champs Magn\'etiques Intenses, EMFL, CNRS UPR 3228, Universit{\'e} Grenoble Alpes, Universit{\'e} Toulouse, Universit{\'e} Toulouse 3, INSA-T, Grenoble and Toulouse, France}
\author{Alessandro Surrente}
\affiliation{Department of Experimental Physics, Faculty of Fundamental Problems of Technology, Wroclaw University of Science and Technology, 50-370 Wroclaw, Poland}

\author{Dimitar Pashov}
\affiliation{King’s College London, Theory and Simulation of Condensed Matter, The Strand, WC2R 2LS London, UK}
\author{Kseniia Mosina}
\affiliation{Department of Inorganic Chemistry, University of Chemistry and Technology Prague, Technicka 5, Prague 6, 16628 Czech Republic}

\author{Zdenek Sofer}
\affiliation{Department of Inorganic Chemistry, University of Chemistry and Technology Prague, Technicka 5, Prague 6, 16628 Czech Republic}

\author{Mark~van Schilfgaarde}
\affiliation{National Laboratory of the Rockies, Golden, CO, USA}

\author{Adam Babiński}
\affiliation{Institute of Experimental Physics, Faculty of Physics, University of Warsaw, Pasteura 5, 02-093, Warsaw, Poland}

\author{Maciej R. Molas}
\affiliation{Institute of Experimental Physics, Faculty of Physics, University of Warsaw, Pasteura 5, 02-093, Warsaw, Poland}

\author{Gabriela Komorowska}
\affiliation{Warsaw University of Technology, Faculty of Materials Science and Engineering, Woloska 141, 02-507 Warsaw, Poland}

\author{Esteban Zamora-Amo}
\affiliation{2D Foundry Research Group, Instituto de Ciencia de Materiales de Madrid (ICMM-CSIC), Madrid E-28049, Spain}

\author{Andres Castellanos-Gomez}
\affiliation{2D Foundry Research Group, Instituto de Ciencia de Materiales de Madrid (ICMM-CSIC), Madrid E-28049, Spain}
\author{Federico Mompeán}
\affiliation{2D Foundry Research Group, Instituto de Ciencia de Materiales de Madrid (ICMM-CSIC), Madrid E-28049, Spain}

\author{Mar Garcia-Hernandez}
\affiliation{2D Foundry Research Group, Instituto de Ciencia de Materiales de Madrid (ICMM-CSIC), Madrid E-28049, Spain}
\author{Micha{\l} Baranowski}\email{michal.baranowski@pwr.edu.pl}
\affiliation{Department of Experimental Physics, Faculty of Fundamental Problems of Technology, Wroclaw University of Science and Technology, 50-370 Wroclaw, Poland}

\author{Swagata~Acharya}\email{swagata.acharya@nlr.gov}
\affiliation{National Laboratory of the Rockies, Golden, CO, USA}

\author{Paulina Plochocka}\email{paulina.plochocka@lncmi.cnrs.fr}
\affiliation{Department of Experimental Physics, Faculty of Fundamental Problems of Technology, Wroclaw University of Science and Technology, 50-370 Wroclaw, Poland}
\affiliation{Laboratoire National des Champs Magn\'etiques Intenses, EMFL, CNRS UPR 3228, Universit{\'e} Grenoble Alpes, Universit{\'e} Toulouse, Universit{\'e} Toulouse 3, INSA-T, Grenoble and Toulouse, France}

\date{\today}

\begin{abstract}
\end{abstract}

\maketitle

\subsection*{Sample characterization}

Figure S1 shows the SEM-EDX characterization of CrSBr$_{1-x}$Cl$_x$ flakes. The elemental maps of the Cl and Br distributions confirm a homogeneous spatial distribution of both halogen species, with no evidence of phase separation or clustering on the micrometre scale.

\begin{figure*}
    \centering
    \includegraphics[width=0.83\textwidth]{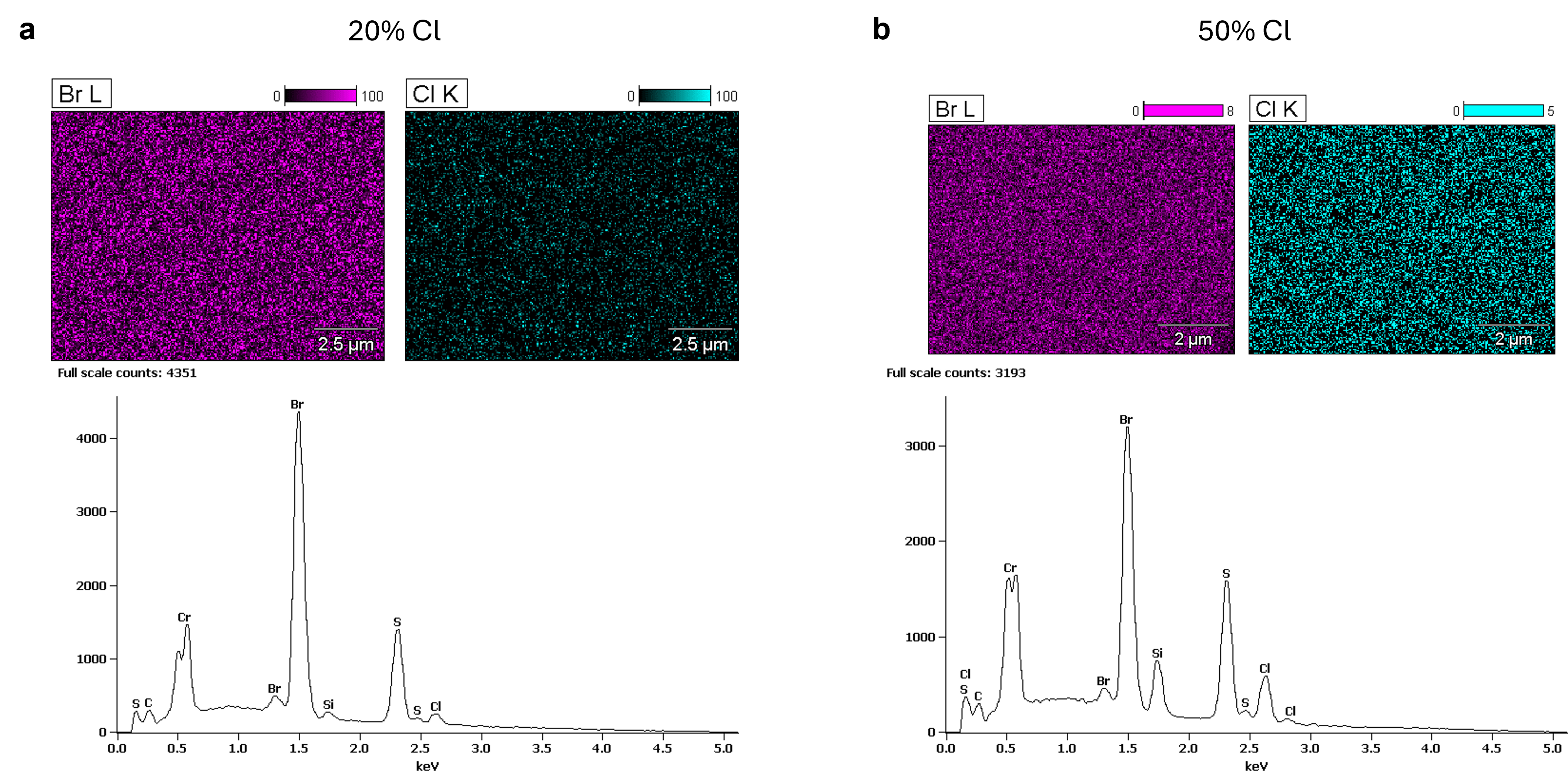}
    \caption{{\textbf{SEM-EDX characterization of thick ($>100$ nm) CrSBr$_{1-x}$Cl$_x$ flakes on a Si/SiO$_2$ substrate.}  Shown are elemental maps of the Br and Cl distributions, characteristic EDX emission spectra as a function of energy, for flakes labeled as 20\% Cl (a) and 50\% Cl (b). }}
    \label{fig:Fig3}
\end{figure*}

\clearpage

\subsection*{Low magnetic field measurements}

Figure~S2 shows the evolution of the reflectance contrast spectra  as a function of magnetic field for flakes of all studied compositions. As in pristine CrSBr, all samples exhibit a characteristic redshift with increasing magnetic field that saturates above a well-defined critical field, reflecting the AFM-to-FM phase transition. The critical field, extracted from the kink in the exciton energy evolution, decreases systematically with increasing Cl content, consistent with the values reported in the main text and confirmed by SQUID magnetometry.
\begin{figure}[H]
    \centering
    \includegraphics[width=0.8\textwidth]{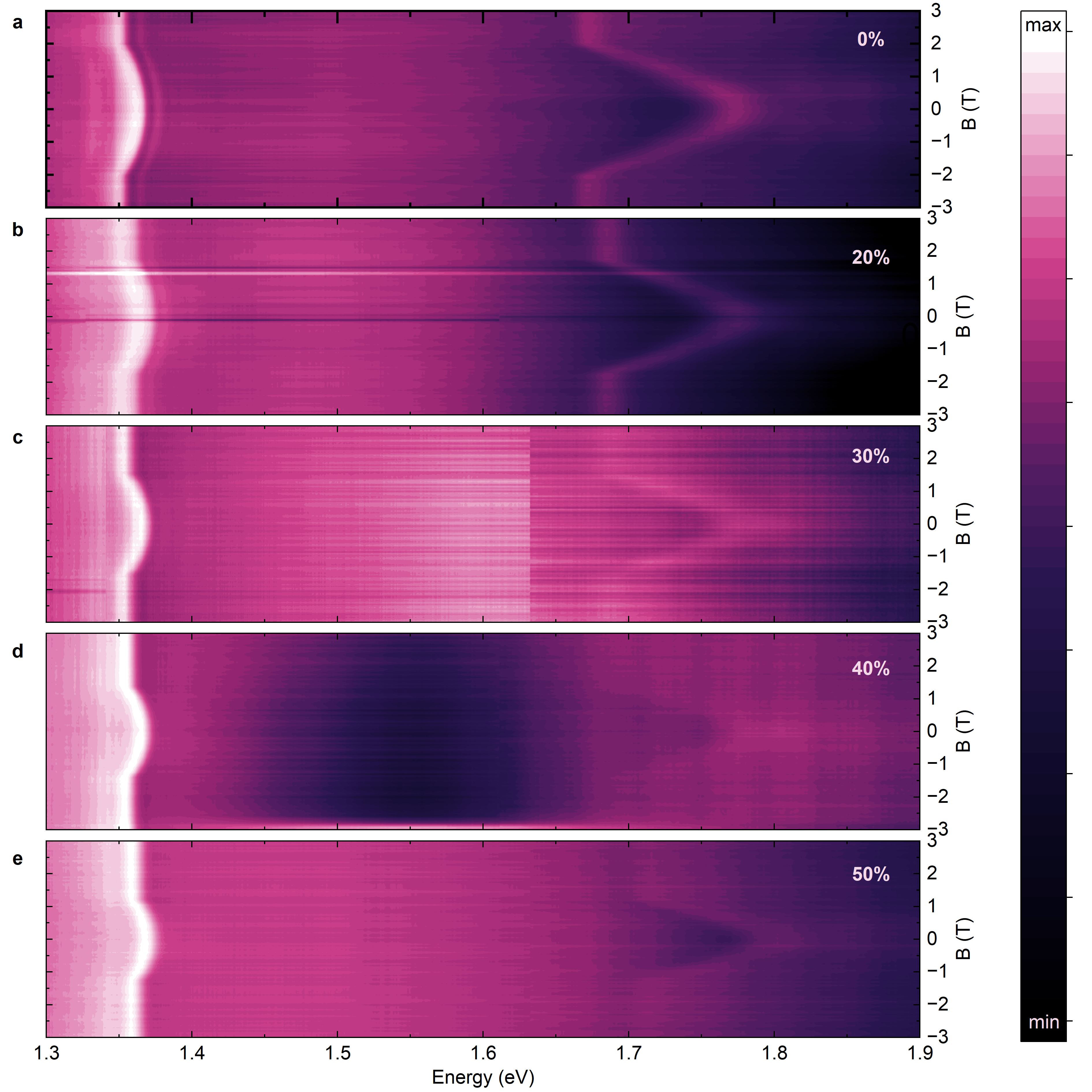}
    \caption{\textbf{Reflectance contrast spectra as a function of magnetic field.} False-colour maps of the reflectance contrast spectra capturing both X$_\text{A}$ and X$_\text{B}$ transition as a function of magnetic field for CrSBr$_{1-x}$Cl$_x$ with Cl content ranging from 0\% to 50\%.}
    \label{fig:S2}
\end{figure}

\clearpage
\subsection*{Wave-functions in AFM phase}

Figure ~S3 shows the real-space exciton wavefunction isosurfaces for the X$_\text{A}$ and X$_\text{B}$ transitions calculated for pristine CrSBr and CrSBr$_{0.33}$Cl$_{0.67}$ in the AFM phase - complementary for the FM phase graphs in the main text. For both excitons and across all Cl compositions in the AFM phase, the wavefunction is confined within a single vdW layer, due to the AFM coupling between neighbouring layers.
The X$_\text{B}$ wavefunction, is substantially reduced in spatial extent, both for AFM and FM phases indicating a progressive evolution toward the Frenkel limit. 

\begin{figure*}[h]
    \centering
    \includegraphics[width=0.8\textwidth]{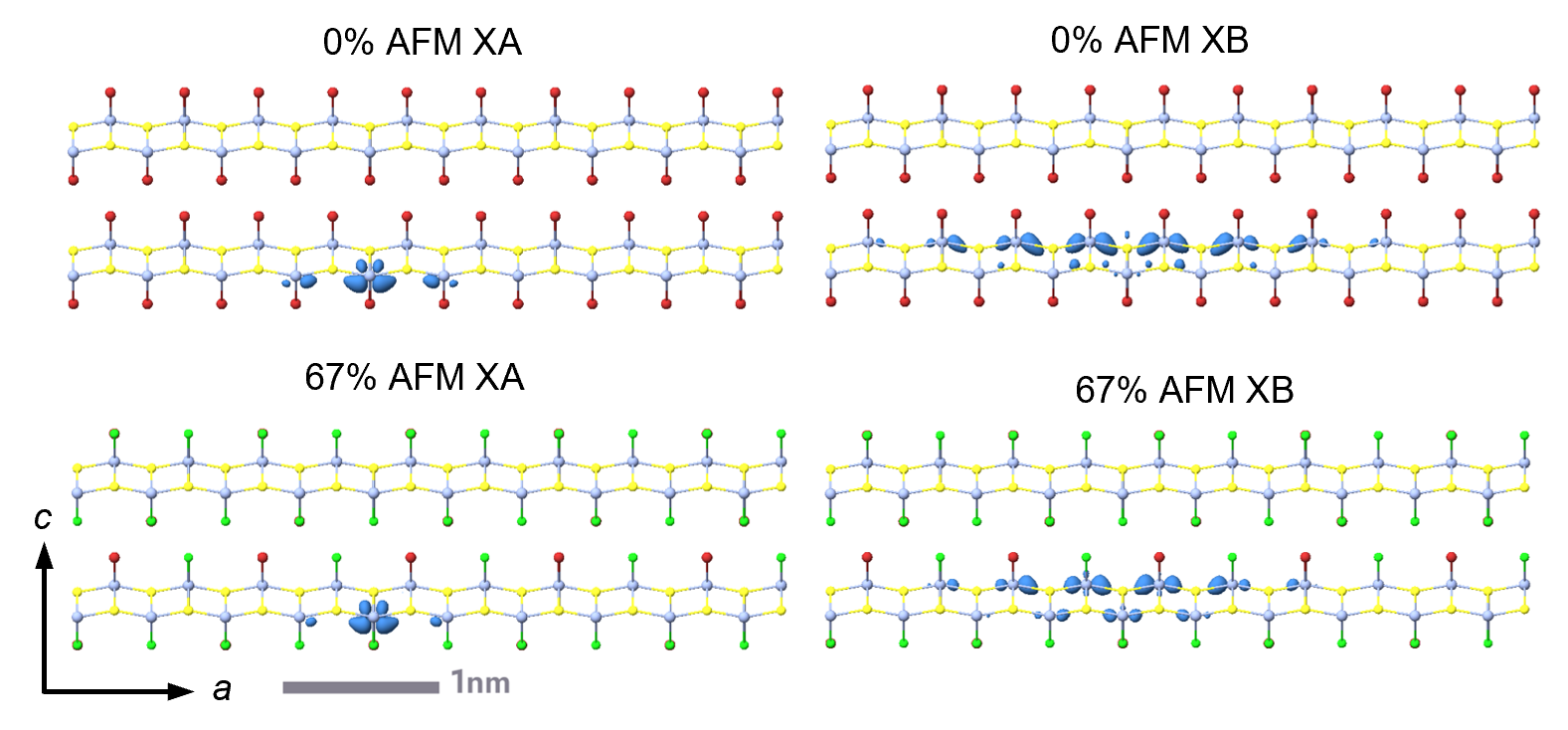}
    \caption{{\textbf{Wavefunction isosurface simulations in AFM phase.} Real-space exciton wavefunction isosurfaces for the X$_\text{A}$ and X$_\text{B}$ transitions in the AFM phase, calculated for pristine CrSBr (0\% Cl, a-b) and CrSBr$_{0.33}$Cl$_{0.67}$ (67\% Cl, c-d). The scale bar corresponds to 1~nm.}}
    \label{fig:S3}
\end{figure*}

\clearpage

\subsection*{High magnetic field measurements}
Figure S4 shows the exemplary magnetic-field-induced energy shifts of the X$_\text{B}$ exciton in the FM phase for all compositions. The data exhibit the characteristic quadratic dependence expected for the diamagnetic shift, following $\Delta E = \sigma B^2$, from which the diamagnetic coefficient $\sigma$ is extracted for each composition. A systematic reduction of $\sigma$ with increasing Cl content is observed across the entire series, providing direct experimental confirmation of the progressive localisation of the X$_\text{B}$ wavefunction upon Cl incorporation, as predicted by the electronic structure calculations and discussed in the main text.

\begin{figure}[h]
    \centering
    \includegraphics[]{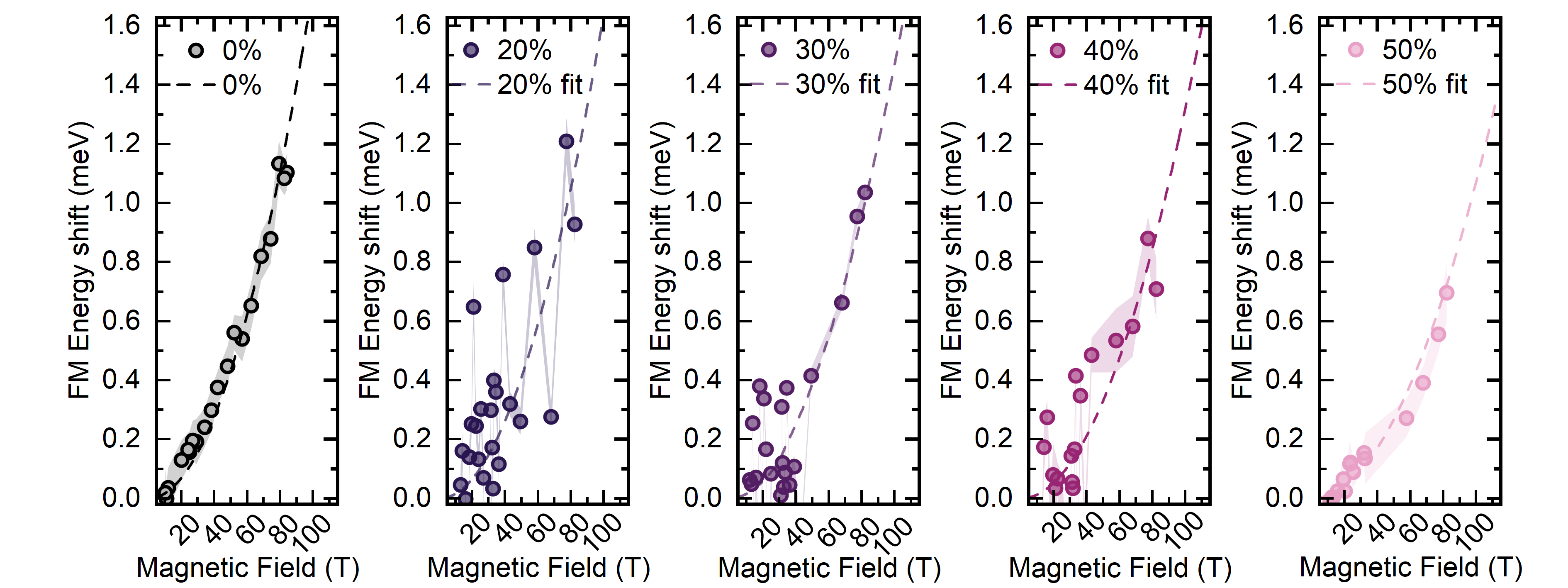}
    \caption{{\textbf{Diamagnetic shifts of the X$_\text{B}$ exciton.} Magnetic-field-induced energy shifts of X$_\text{B}$ in the FM phase for CrSBr$_{1-x}$Cl$_x$ with Cl content ranging from 0\% to 50\%. Dashed lines are parabolic fits used to extract the diamagnetic coefficients $\sigma$.}}
    \label{fig:Fig3}
\end{figure}

\clearpage

\bibliography{Bibliography_Experiment,Bibliography_Theory}